\documentclass[aps,prb, twocolumn, showpacs, superscriptaddress]{revtex4-1}
\usepackage[utf8]{inputenc}
\usepackage{graphicx}
\usepackage{pstricks}
\usepackage{array}
\usepackage{natbib}
\usepackage{amsmath}
\usepackage{pifont}
\usepackage{dsfont}
\usepackage{multirow}
\usepackage{amssymb}
\usepackage{wasysym}
\usepackage{breqn}

\def\be{\begin{equation}}
\def\ee{\end{equation}}

\begin{document}

\title{A linear-scaling source-sink algorithm for simulating time-resolved quantum transport and superconductivity}

\newcommand{\spsmsA}{Univ. Grenoble Alpes, INAC-SPSMS, F-38000 Grenoble, France}
\newcommand{\spsmsB}{CEA, INAC-SPSMS, F-38000 Grenoble, France}

\newcommand{\xbar}{\ensuremath{\bar{x}}}
\newcommand{\SigmaBar}{\ensuremath{\bar{\Sigma}}}
\newcommand{\PsiBar}{\ensuremath{\bar{\Psi}}}
\newcommand{\phiBar}{\ensuremath{\bar{\phi}}}
\newcommand{\order}[1]{\ensuremath{\mathcal{O}\left( #1 \right)}}

\author{Joseph Weston}
\affiliation{\spsmsA}
\affiliation{\spsmsB}
\author{Xavier Waintal}
\affiliation{\spsmsA}
\affiliation{\spsmsB}
\date{\today}

\begin{abstract}
We report on a ``source-sink'' algorithm which allows one to calculate time-resolved physical
quantities from a general nanoelectronic quantum system (described by an arbitrary time-dependent quadratic Hamiltonian) connected to infinite electrodes. Although mathematically equivalent to the non equilibrium Green's function formalism, the approach is based on the scattering wave functions of the system. It amounts to solving a set of generalized Schrödinger equations which include an additional ``source'' term (coming from the time dependent perturbation) and an absorbing ``sink'' term (the electrodes). The algorithm execution time scales linearly with both system size and simulation time allowing one to simulate large systems (currently around $10^6$ degrees of freedom) and/or large times (currently around $10^5$ times the smallest time scale of the system). As an application we calculate the current-voltage characteristics of a
Josephson junction for both  short and long junctions, and recover the multiple Andreev reflexion (MAR) physics. We also discuss two intrinsically time-dependent situations:  the relaxation time of a Josephson junction after a quench of the voltage bias, and the propagation of voltage pulses through a Josephson junction. In the case of a ballistic, long Josephson junction, we predict that a fast voltage pulse creates an oscillatory current whose frequency is controlled by the Thouless energy of the normal part. A similar effect is found for short junctions; a voltage pulse produces an oscillating current which, in the absence of electromagnetic environment, does not relax.
\end{abstract}

\maketitle
As quantum nanoelectronics experiments get faster (in the GHz range and above) it
becomes possible to study the time dependent dynamics of devices in their quantum
regimes, i.e. at frequencies higher than the system temperature ($1$K corresponds roughly to $20$GHz). Recent achievements include coherent single electron sources with well defined release time\cite{Glattli2013} or energy\cite{feve_on-demand_2007}, pulse propagation along quantum Hall edge states\cite{Ashoori1992,Kamata2010,Fujisawa2011} and
terahertz measurements in carbon nanotubes\cite{zhong_terahertz_2008}. While the mathematical framework for describing
quantum transport in the time domain has been around since the 90s \cite{wingreen_time-dependent_1993,jauho_time-dependent_1994}, the
corresponding non-equilibrium Green's function formalism (NEGF) is rather cumbersome and can only be
solved in rather simple situations, even with the help of numerics. In Ref.~\onlinecite{Twave_formalism} we developed an alternative formulation of the theory which is much easier to solve numerically, in addition to being more physically transparent. The approach of Ref.~\onlinecite{Twave_formalism} (to which we refer for further references) was recently used in a variety of situations including electronic interferometers\cite{fabryperot,gaury_josephson_2015}, quantum Hall effect\cite{gaury_stopping_2014}, normal-superconducting junctions\cite{weston_manipulating_2015}, Floquet topological insulators\cite{fruchart_probing_2016} and the calculation of the quantum noise of voltage pulses\cite{gaury_computational_2016}.

The best algorithm introduced in Ref.~\onlinecite{Twave_formalism} (nicknamed WF-C)
has a computational execution time that scales linearly with the system size $N$, but as the square of the total simulation time. While for ballistic systems this $t_{max}^2$ limitation was not too stringent, in situations with large separations of time scales (such as the Josephson junctions studied below), it makes the numerical calculation computationally prohibitive.
In this manuscript, we present an extension of the previous approach which reduces the computational
complexity down to $\order{N t_{max}}$. This is achieved with the addition of non-hermitian terms, referred to as ``sink'' terms, in the Hamiltonian in addition to the ``source'' terms introduced in the WF-C method of Ref.~\onlinecite{Twave_formalism}. The new technique remains mathematically equivalent to the NEGF formalism.

This articles is organized as follows. Section \ref{sec:model} introduces a general class of models and the time dependent scattering states of the system. In section \ref{sec:source} we briefly recall how a simple change of variables leads to the introduction of an additional source term in the Schrödinger equation, which greatly facilitates the numerical treatment. In section \ref{sec:sink} we develop the
new part of the algorithm and show how the introduction of sink terms solves previous difficulties
at long times. Finally, section \ref{sec:jj} discusses applications to the physics of out of equilibrium Josephson
junctions. After recovering well known effects (Multiple Andreev Reflection in both short and long junctions, AC Josephson effect, relaxation of Andreev bound states), we study the propagation of fast voltage pulses through
Josephson junctions.

\section{Model}
\label{sec:model}

\begin{figure}[b]
    \includegraphics[width=0.47\textwidth]{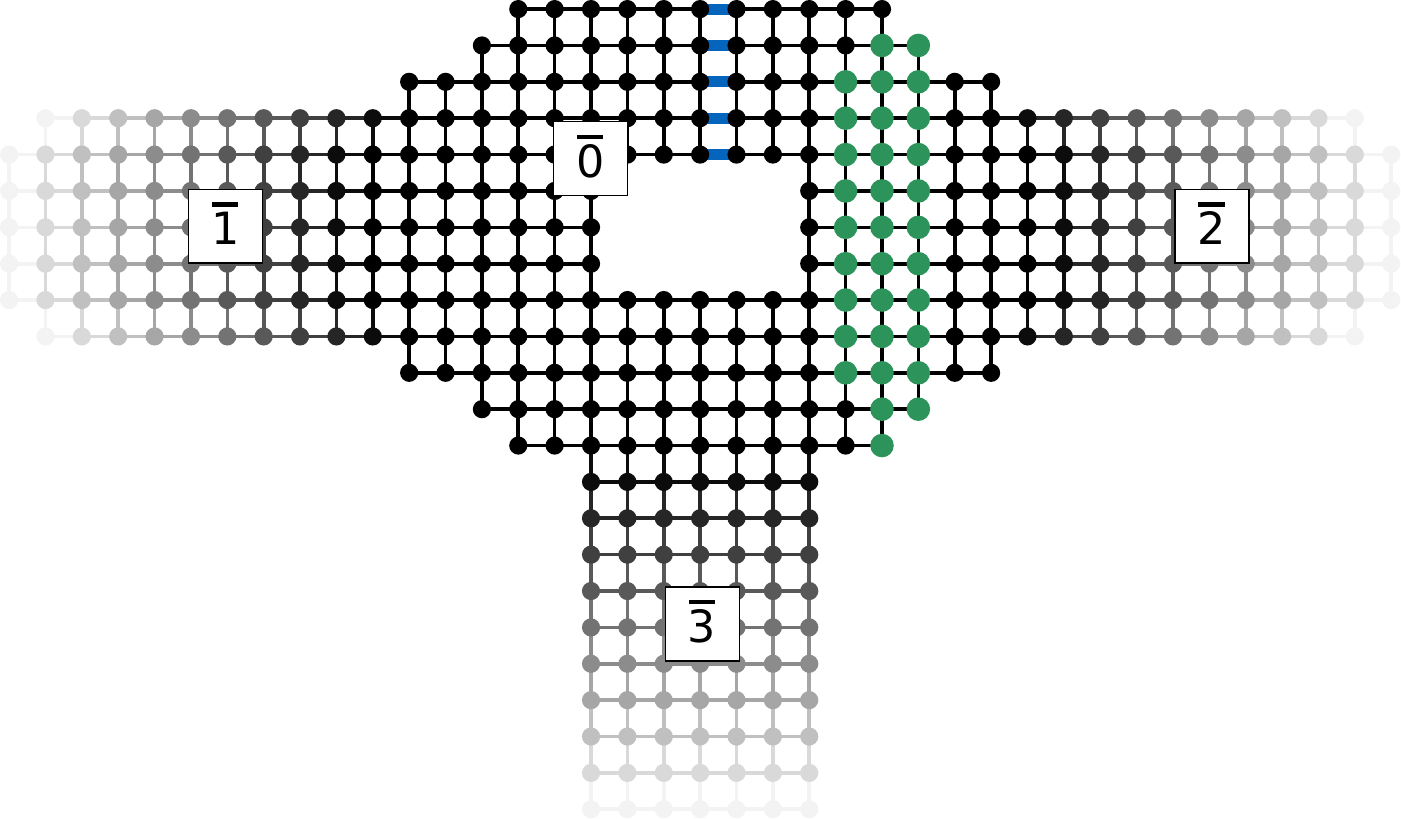}
    \caption{Sketch of a typical system considered. It consists of a central scattering region, $\bar{0}$,
        attached to semi-infinite leads $\bar{1}$, $\bar{2}$, and $\bar{3}$. Some of the on site potentials
        are time dependent (for instance the bold green sites correspond to the sites unerneath a pulsed
        electrostatic gate) as well as some of the inter site hoppings (for instance the blue connections correspond to
        a time dependent magnetic field sent through the central hole).}
    \label{fig:system}
\end{figure}

We consider a general class of models describing a quantum device of finite extent
attached to semi-infinite electrodes. The full system is described by a general quadratic Hamiltonian of the form
\begin{equation}
    \mathrm{\hat H}(t) =
    \sum_{ij} \mathrm{H}_{ij}(t) \hat{c}^{\dagger}_{i}\hat{c}_{j}
    \label{eq:ham}
\end{equation}
where $\hat{c}^{\dagger}_i$ ($\hat{c}_j$) are the Fermionic creation (annihilation)
operators of a one-particle state on site $i$. A ``site'' $i$ typically labels position
as well as other degrees of freedom such as spin, orbital angular momentum or electron/hole
(as in the superconducting application below). The $\mathrm{H}_{ij}(t)$ are the matrix elements
of the Hamiltonian matrix $\mathbf{H}(t)$.
The system consists of a time-dependent central region $\bar 0$ connected to several leads $\bar 1,\bar
2,...$ as depicted in Fig.~\ref{fig:system}. We keep the Hamiltonian of the
central region fully general but restrict the leads to be semi-infinite, time
independent and invariant by translation (i.e. they have a quasi-one dimensional periodic
structure). Each lead remains in its thermal equilibrium at all times. We further
suppose that the time-dependent perturbations are only switched on at positive
times, so that $\mathbf{H}(t<0) =\mathbf{H}_0$. Note that
if one has a uniform time-varying potential in one or more of the leads then
a gauge transformation can always be performed such that the time-dependence
is brought into the interface between the lead and the central region, which
can then be included in the definition of the central region. Typically the time dependent
part of the Hamiltonian is restricted to rather small regions as illustrated in Fig.~\ref{fig:system}.

Before the time-dependent perturbations are switched on, the system is
characterized by its scattering wavefunctions $\Psi_{\alpha E}$ that are labeled by
their energy $E$ and incoming channel $\alpha$,
\be
\mathrm{\textbf{H}}_0
\Psi^{st}_{\alpha E}= E \Psi^{st}_{\alpha E}.
\ee
The scattering states $\Psi^{st}_{\alpha E}$ are standard object of mesoscopic physics
and can be obtained directly by wave matching the incoming and outgoing modes
at the lead-system boundary. For complicated
geometries these can be obtained numerically by using e.g. the Kwant package\cite{Kwant_preparation}.
A physical observable
$\mathrm{\hat{A}} = \sum_{ij} \mathrm{A}_{ij}
\hat{c}^{\dagger}_{i}\hat{c}_{j}, $ (e.g. electronic density or local currents) can be directly obtained from the knowledge of these wavefunctions by simply filling up the one-body scattering states according to Fermi statistics, using
\be
\langle \mathrm{\hat{A}} \rangle =
\sum_{\alpha}\int \frac{dE}{2\pi} f_\alpha(E) \Psi^{st\,\dagger}_{\alpha E}
\mathrm{\textbf{A}} \Psi^{st}_{\alpha E} \label{eq:st}
\ee
where $f_\alpha (E)$ is the Fermi function of the electrode associated with channel $\alpha$.
The celebrated Landauer formula for the conductance is a special case of Eq.(\ref{eq:st}).

The generalization of Eq.(\ref{eq:st}) to the time-dependent problem is rather
straightforward: one first obtains the scattering states and lets them evolve
according to the Schrödinger equation
\be
\label{eq:pfree}
i \partial_{t} \Psi_{\alpha E}(t) = \mathrm{\textbf{H}} (t) \Psi_{\alpha E}(t)
\ee
with the initial condition $\Psi_{\alpha E}(t=0)=\Psi^{st}_{\alpha E}$. The
observables follow from Eq.(\ref{eq:st}) where the $\Psi^{st}_{\alpha E}$ are
replaced by $\Psi_{\alpha E}(t)$:
\be
\label{eq:time-dep-obs}
\langle \mathrm{\hat{A}}(t) \rangle =
\sum_{\alpha}\int \frac{dE}{2\pi} f_\alpha(E) \Psi_{\alpha E}^\dagger(t)
\mathrm{\textbf{A}} \Psi_{\alpha E}(t)
\ee
The fact that such a scheme is equivalent to
the NEGF formalism or to the scattering approach was derived
in Ref.~\onlinecite{Twave_formalism}. In particular, the central objects of the NEGF
formalism, the so-called lesser ($<$), greater ($>$) and retarded ($R$) Green's functions, have simple expressions in term of the time dependent scattering states,
\begin{eqnarray}
    G_{ij}^<(t,t') &\equiv&  i \langle \hat{c}^\dagger_j(t') \hat{c}_i(t) \rangle\\
&=&\sum_\alpha \int \frac{dE}{2\pi}\ if_\alpha(E)  \Psi_{\alpha E}(t,i)
\Psi_{\alpha E}^*(t',j)\nonumber
\end{eqnarray}
\begin{eqnarray}
G_{ij}^>(t,t') &\equiv&  -i \langle \hat{c}_i(t) \hat{c}^\dagger_j(t')  \rangle\\
&=&\sum_\alpha \int \frac{dE}{2\pi}\ i[f_\alpha(E) - 1]  \Psi_{\alpha E}(t,i)
\Psi_{\alpha E}^*(t',j)\nonumber
\end{eqnarray}
\begin{eqnarray}
G_{ij}^R(t,t') &\equiv&  -i \theta(t-t')\langle \hat{c}^\dagger_j(t') \hat{c}_i(t) + \hat{c}_i(t) \hat{c}^\dagger_j(t') \rangle\\
&=&-i\theta(t-t') \sum_\alpha \int \frac{dE}{2\pi}\   \Psi_{\alpha E}(t,i)
\Psi_{\alpha E}^*(t',j)\nonumber
\end{eqnarray}
Note that in the presence of bound states (such as the Andreev states in the Josephson junctions described below) the above integral need to be replaced by an integral over the continuum plus a sum over the bound states, as explained in Ref.~\onlinecite{profumo_quantum_2015}.

\section{The source}
\label{sec:source}

In its original form, Eq.(\ref{eq:pfree}) is not very
useful for numerics because the wave function spreads over the entire {\it
infinite} system. A first simple, yet crucial, step consists in introducing the deviation from the
stationary solution, $\bar\Psi_{\alpha E}(t)$,
\be
    {\Psi}_{\alpha E}(t) = e^{-iEt}(\Psi_{\alpha E}^{st} + \bar\Psi_{\alpha E}(t)).
\label{Psibar}
\ee
${\bar \Psi}_{\alpha E}(t)$ satisfies,
\be
\label{eq:wbl}
    i \partial_{t} \bar\Psi_{\alpha E}(t) =
    [\mathrm{\textbf{H}}(t) - E]\bar\Psi_{\alpha E}(t) + S_{\alpha E}(t),
\ee
with
\be
    S_{\alpha E}(t) =
    [\mathrm{\textbf{H}}(t) - \mathrm{\textbf{H}}_0]
    \Psi_{\alpha E}^{st}
\ee
and
\be
\bar\Psi_{\alpha E}(t=0)=0.
\ee
The new ``source'' term $S_{\alpha E}(t)$ can be computed from the
knowledge of the stationary scattering states and is localized at the place where the
time-dependent perturbation takes place (where $\mathrm{\textbf{H}}(t) \ne \mathrm{\textbf{H}}_0$,
typically the colored regions of Fig.~\ref{fig:system}).
Eq.~\eqref{eq:wbl} is already much better than Eq.~\eqref{eq:pfree} for numerics because the
initial condition corresponds to a wavefunction that vanishes everywhere. One
can therefore truncate Eq.~\eqref{eq:wbl} and keep a finite system around the
central time-dependent region where the source term lies.  In practice, one adds $N$ layers
of each electrodes. Note that in order for this procedure to be correct, the stationary scattering states are calculated for the {\it infinite} system and the truncation is only performed afterwards. For the truncation to be valid, the size of this finite region must be larger than $N > v\,t_{max}/2$ where $v$
is the maximum group velocity at which the wavefunction can propagate and $t_{max}$ the duration of the simulation.  Hence,
for large values of $t_{max}$, the total computational time to integrate
Eq.~\eqref{eq:pfree} scales as $v\,t_{max}^2$. This algorithm corresponds to the WF-C
algorithm of Ref.~\onlinecite{Twave_formalism}.
Here, we have explicitly removed a factor $e^{-iEt}$ from the definition of
$\bar\Psi_{\alpha E}(t)$ compared to Ref.~\onlinecite{Twave_formalism}. This
change, while small, leads to an improve stability of the numerical
integration: the equation of motion for $\bar\Psi_{\alpha E}(t)$ does not have
an (potentially fast) oscillating factor $e^{-iEt}$ in the source term. This
means that the numerical integration scheme used for solving eq.~(\ref{eq:wbl})
is now limited by the intrinsic timescales of the problem, and not the
``artificial'' timescale $\hbar/E$ introduced by a bad choice of gauge.

\section{The sink}
\label{sec:sink}

The $t_{max}^2$ scaling of the algorithm comes from the fact that for long simulation times, one needs
to introduce large part of the leads ($\propto t_{max}$) in order to avoid spurious reflections at the boundaries where the leads have been truncated.
To proceed, one needs to take advantage of the special structure of the leads: they are not only time-independent, but also invariant by translation. Hence whatever enters into the lead will propagate toward infinity and never come back to the central region. Mathematically, the form of $\bar\Psi_{\alpha E}(t)$ in the leads is a superposition of {\it outgoing} plane
waves~\cite{Twave_formalism}
\be
\label{eq:leads}
\bar\Psi_{\alpha E}(t) = \int \frac{dE'}{2\pi} S_{\alpha' \alpha}(E',E) e^{-iE't+k'n} \xi_{\alpha'}(E')
\ee
where $E'$ and $k'$ are related by the dispersion relation of the lead, $n$
indexes the different unit cells in the lead, $\xi_{\alpha'}$ the transverse
wavefunction of the corresponding mode and $S_{\alpha'\alpha}(E',E)$ is the
time-dependent part of the inelastic scattering matrix. The crucial point of
Eq.(\ref{eq:leads}) is that it only contains outgoing modes as the incoming one
has been subtracted when removing the stationary scattering state.
Therefore, once the wave function starts to reach the leads, it propagates
toward infinity and never comes back to the central system.

A natural idea that comes to mind is to replace the finite fraction of the electrodes
by some sort of (non-hermitian) term in the Hamiltonian that ``absorbs'' the wavefunction
that enters the leads. This has been studied in the literature in the context of various partial differential equations,\cite{Antoine2008, muga_complex_2004,shemer_optimal_2005, riss_transformative_1998, riss_reflection-free_1995, kalita_use_2011, ge_use_1998} and
is usually known as a complex absorbing potential. The difficulty lies in the fact that this
absorbing term must not give rise to reflections. At a given energy, a
perfectly absorbing boundary condition does exist, it corresponds to adding the self energy
of the lead at the boundary (which is a non-local complex absorbing potential, see WF-D method
of Ref.~\onlinecite{Twave_formalism}). However the outgoing waves of Eq.(\ref{eq:leads}) span a finite energy window so that some energies would get reflected back to the central region. One solution to obtain a perfectly absorbing boundary condition is
to use a boundary condition that is non local in time\cite{Antoine2008}, as in the WF-B method of
Ref.~\onlinecite{Twave_formalism}; this leads to algorithms that scale as $t_{max}^2$.

\begin{figure}[t]
    \includegraphics[width=0.47\textwidth]{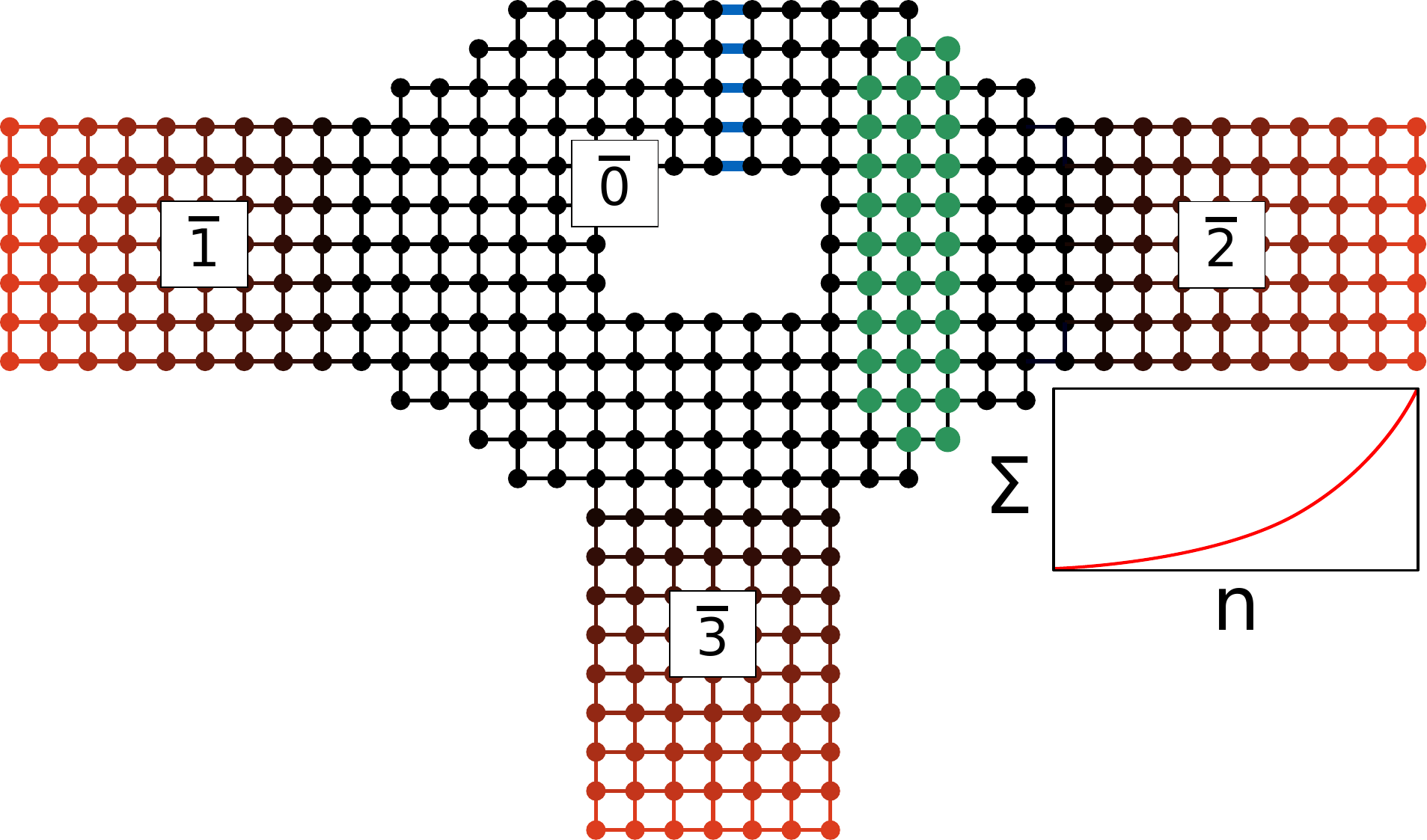}
    \caption{Sketch of the truncated approximation to the system shown in Figure~\ref{fig:system}, including the absorbing layers.
        The (red) color of the sites indicates the intensity of the complex absorbing
    potential. The curve next to lead 2's absorbing layer
    shows a typical shape of the complex absorbing potential, $\Sigma$.}
    \label{fig:extended-system}
\end{figure}

We choose instead to design an imaginary potential that varies spatially. We show
that {\it for any desired accuracy}, we can design an imaginary potential that spreads over
a finite width of $N$ electrode unit cells - where $N$ depends only of the required accuracy, not on $t_{max}$. In practice, this new algorithm is much more effective than WF-C when $t_{max}$ becomes larger than the ballistic time of flight through the system. The idea behind the algorithm is fairly straightforward: suppose that a plane wave with a dispersion relation $E(k)$ propagates inside one electrode. If one adds an imaginary potential $-i\Sigma$ to the Schrödinger equation, this plane wave becomes evanescent which eventually leads to the absorption of the wave.
On the other hand, any abrupt variation of potential (or in this case of imaginary potential) leads
to unwanted reflection back to the central part of the system. Hence, the algorithm consists in adiabatically switching on the imaginary potential $\Sigma (n)$ inside a finite fraction of the electrode, see  Fig.~\ref{fig:extended-system} for a sketch.
The new equation of motion contains both the previous source term and the additional sink in the electrodes,
\be
    i \partial_{t} \bar\Psi_{\alpha E}(t) =
    \left[\mathrm{\textbf{H}}(t) - E -
    i\mathrm{\bf \Sigma }
    \right]\bar\Psi_{\alpha E}(t) +
    S_{\alpha E}(t),
\ee
where the matrix $\mathrm{\bf \Sigma }$ is diagonal and vanished in the central region while it reads
\be
\mathrm{\bf \Sigma } = \Sigma (n)\ \mathbf{1}_{cell}
\ee
in the absorbing layer placed at the beginning of the electrodes. The index $n$ labels the unit cells of the leads and $\mathbf{1}_{cell}$ is the identity matrix defined over a unit cell. What remains to be done is
 to specify the function $\Sigma (n)$ so that it is large enough to absorb all waves entering into
 the lead while being smooth enough not to produce spurious reflections. The error induced by the boundary
 conditions must not exceed a tolerance $\delta$. Our aim is to minimize the number $N$ of layers that must be added in the simulation to absorb the outgoing waves without the error exceeding $\delta$.

\subsection{Analytical Calculation of the spurious reflection}
Before we can design a suitable imaginary potential, we must understand how the spurious reflection
back to the central part depends on the shape of $\Sigma (n)$.
We will start from a continuum model in order to develop an analytical solution for this simple case.
The rationale, other than its tractability, is the fact that spurious reflections happen
when $\Sigma (n)$ varies on a spatial scale that is short  {\it compared to
the wavelength } of the solution, hence is dominated by small momentum $k$ where the tight-binding
dispersion relation reduces to its continuum limit. We will show that there is an extremely good agreement
between the analytical results derived in this section and numerical calculations of the discretized model.

Let us consider the stationary 1D Schrödinger equation,
\be
\label{eq:se}
    -\frac{\hbar^2}{2m^*} \frac{\partial^2 \psi(x)}{\partial x^2} -
        \frac{i}{L}\Sigma\left(\frac{x}{L}\right)\psi(x) = E\psi(x)
\ee
where $m^*$ is the electron effective mass and we have introduced a length
scale, $L$, which controls how fast $\Sigma(x)$ varies. For negative $x$, we set
$\Sigma(x\le0) = 0$ so that the wave function is in a superposition of plane waves,
\be
    \psi(x) = e^{i kx} + r_\Sigma e^{-i kx}
\ee
where we \emph{define} $E = \hbar^2k^2/2m^*$. Our goal is to calculate the spurious reflection probability $R_\Sigma = |r_\Sigma|^2$ induced by the presence of the imaginary potential.
We first rescale the equation by $E$ and define $\xbar = kx$, $\SigmaBar(u) = (k/E)\Sigma(u)$
and $\psi(x) = \bar\psi(\xbar)$ to obtain the dimensionless equation,
\be
\label{eq:se-scale}
    \left[\partial^2_{\xbar}
          + \frac{i}{kL}\SigmaBar\left(\frac{\xbar}{kL}\right)
          + 1
    \right]
    \bar\psi(\xbar) = 0
\ee
with
\be
\label{eq:bc}
    \bar\psi(\xbar) = e^{i\xbar} + r_\Sigma e^{-i\xbar}
\ee
for $\xbar <0$. It is apparent from Eq.(\ref{eq:se-scale}) that the spurious reflection
is controlled by the dimensionless parameter $kL$. Since we want this spurious reflection to be small,
we will work in the limit of large $kL\gg 1$ and expand $r_\Sigma$ in powers of $1/kL$.
The zeroth order contribution is simply the extension of the WKB limit to imaginary potential; the
wave function takes the form of an evanescent wave,
\be
\bar \psi(\bar x) \approx e^{\bar S(\xbar)}
\ee
with $\bar S(\xbar)$ satisfying
\be
    \label{eq:WKB_exp}
    [\bar{S}'(\xbar)]^2 + 1 + i\frac{1}{kL}\SigmaBar(\frac{\xbar}{kL})= 0
\ee
where primes denote derivatives. We expand $\bar S(\xbar)$ to first order
in $1/kL$, and apply the boundary condition Eq.~\eqref{eq:bc}
at $\xbar=0$, as well as $\bar \psi(kL) = 0$ (perfect reflection at a the end of
the simulation domain at $x=L$) to obtain the zeroth order contribution to
$r_\Sigma$:
\be
    \label{eq:r0}
    r^0_\Sigma = e^{2ikL}e^{-Ak/E},
\ee
where
\be
    A = \int_0^L \frac{1}{L}\Sigma\left(\frac{x}{L}\right)\, dx
\ee
is independent of $kL$.  Physically speaking, the wave function is
exponentially attenuated up to the hard wall at $x=L$ where it is fully
reflected and then again exponentially attenuated until $x=0$.

The contribution $r^0_\Sigma$ takes into account the finite absorption due to the
imaginary potential but not the spurious reflections due to wavevector mismatch.
It it therefore necessary to go beyond the {\it adiabatic} WKB approximation and calculate
its $1/kL$ deviation $r^1_\Sigma$. We can ignore the hard wall at $x = L$ as it will play no
role in what follows. Generalizing the WKB approximation we choose the following
ansatz for $\xbar > 0$:
\be
\label{eq:sol}
    \bar\psi(\xbar) = \phiBar(\xbar)e^{\bar{S}(\xbar)}
\ee
$\bar{S}(\xbar)$ contains the fast oscillating and decaying parts,
while $\phiBar(\xbar)$ contains the remaining (slow) parts.
Plugging the ansatz Eq.~(\ref{eq:sol}) into Eq.~(\ref{eq:se-scale}) our
Schrödinger equation becomes
\be
\begin{aligned}
\label{eq:new-se}
   \biggl\{
    \phiBar''(\xbar)
    &+\left[ 2i -
            \frac{1}{kL} \SigmaBar\left(\frac{\xbar}{kL}\right) +
            2\order{\frac{1}{(kL)^2}}
    \right]\phiBar'(\xbar)
    \\
    &+\left[
        \frac{-1}{2(kL)^2} \SigmaBar'\left(\frac{\xbar}{kL}\right) +
            \order{\frac{1}{(kL)^3}}
    \right] \phiBar(\xbar)
    \biggr\}
    e^{\bar{S}(\xbar)} = 0
\end{aligned}
\ee
with
\be
    \bar{S}(\xbar) =
        i\xbar -
        \frac{1}{2}\int_0^{\xbar/kL} \SigmaBar(u)\, du +
        \order{\frac{1}{kL}}
\ee
We write $\phiBar(\xbar)$ as
$\phiBar(\xbar) = \phiBar_0(\xbar) + (1/kL)\phiBar_1(\xbar)$
and notice that, in the limit $(1/kL) \to 0$, Eq.~(\ref{eq:new-se})
admits a solution
$\phiBar(\xbar) = \phiBar_0(\xbar) = A + Be^{-2i\xbar}$.
In this limit there should be no backscattering from the
imaginary potential, so $B=0$ and $\phiBar_0(\xbar) = 1$, to match the boundary conditions
Eq.~(\ref{eq:bc}). The derivatives of $\phiBar_0(\xbar)$ hence vanish
and we arrive at
\be
\begin{aligned}
    \phiBar\,''_1(\xbar) +
    2\left[ i - \frac{1}{2kL} \SigmaBar(\xbar/kL)
    \right]\phiBar_1'(\xbar)
    = \frac{1}{2kL} \SigmaBar'(\xbar/kL)
    \label{eq:phi1}
\end{aligned}
\ee
up to terms of order $\order{(1/kL)^2}$. Eq.(\ref{eq:phi1})
can be solved by the variation of constant method,
\be
    \phiBar\,'_1(\xbar) =
        \bar{C}(\xbar)
            \exp{\left[
                -2i\xbar +
                \int_0^{\xbar/kL}\SigmaBar(u)\, du
            \right]}
\ee
with
\be
\label{eq:Aprime}
     \bar{C}\,' (\xbar) =
        \frac{1}{2kL}\SigmaBar'(\xbar/kL)
        \exp{\left[
            2i\xbar -
            \int_0^{\xbar/kL}\SigmaBar(u)\, du
        \right]}
\ee
Applying the continuity condition on $\bar\psi(\xbar)$ and
$\bar\psi'(\xbar)$ at $\xbar=0$ we obtain the $1^{st}$ order
contribution to the reflection amplitude:
\be
    r^1_\Sigma = \frac{-1}{2ikL}\bar{C}(0)
\ee
which we can write explicitly, using Eq.~(\ref{eq:Aprime})
and the condition $\bar{C}(\infty) = 0$, as
\be
\label{eq:r1}
    r^1_\Sigma = \frac{1}{4ikL}\int_0^\infty
            \bar{\Sigma}'(u)
            \exp{\left[2ikLu- \int_0^{u}\bar{\Sigma}(v)\, dv \right]}
        du
\ee
One can understand $r^1_\Sigma$ as the Fourier transform at (large) frequency
($kL$) of the gradient of the imaginary potential
weighted by the absorption that has already taken place.  Putting together Eq.~(\ref{eq:r0}) and Eq.~(\ref{eq:r1}), we finally obtain
\be
\label{eq:reflection}
\begin{aligned}
    r_\Sigma =&\ e^{2ikL}e^{-Ak/E}\\
    &+\frac{1}{4iEL}\int_0^\infty
    \Sigma'(u)
    \exp{\left[
        2ikLu -
        \frac{k}{E}\int_0^{u}\Sigma(v)\, dv
    \right]}
    du.
\end{aligned}
\ee
Eq.(\ref{eq:reflection}) is the main result of this section. Now that we understand
how the spurious reflection depends on the shape of $\Sigma (x)$, we need to design the
imaginary potential so as to minimize Eq.(\ref{eq:reflection}) (for a given $L$).
More precisely, for a given required precision $\epsilon$, we wish to enforce $R_\Sigma < \epsilon$
{\it irrespective } of the value of the energy $E$. Such a stringent condition is not, strictly
speaking, feasible as $R_\Sigma \to 1$ when $E\to 0$ (all the variations of the imaginary potential
become ``abrupt'' when the electronic wave length becomes infinite) but we will see that the associated
error can be kept under control.

\begin{figure*}[t]
    \includegraphics[width=\textwidth]{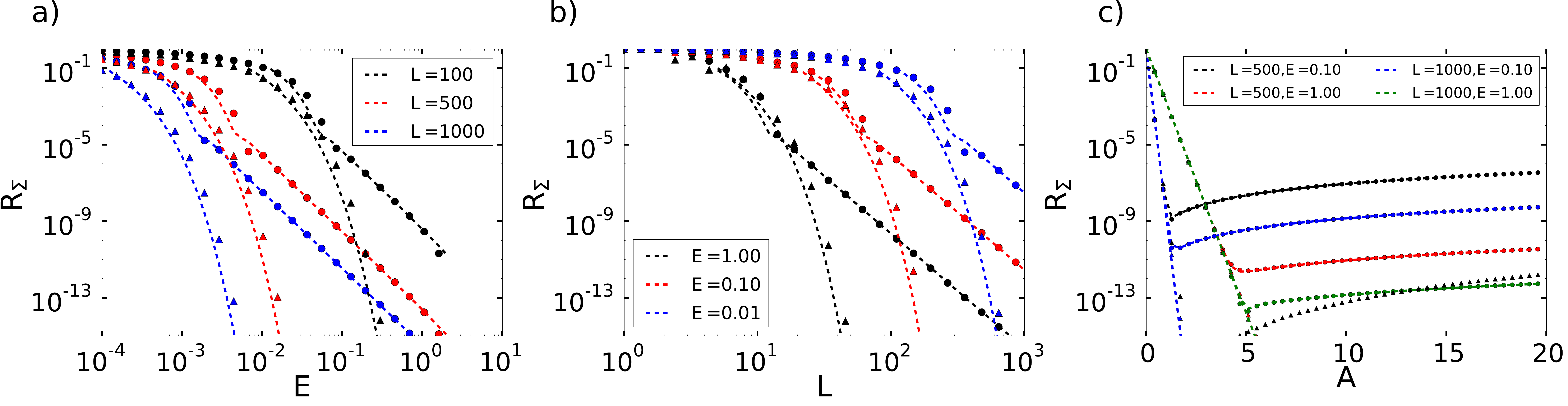}
    \caption{d.c. reflection probability of a one dimensional chain in presence of an imaginary potential. The three panels show the scaling with a) energy,
    b) absorbing region length, and c) area under the imaginary potential curve.
    Symbols are numerical simulation of the discrete model and dashed lines are
    the analytic (continuum) result, Eq.~(\ref{eq:mono-CAP}),
    with $n=2$ (circles) and $n=6$ (triangles).}
    \label{fig:scaling}
\end{figure*}

A shape that keeps $R_\Sigma$ small must be initially very flat and later (when a significant
fraction of the wave has been already absorbed) can increase more rapidly.
We leave a full optimization of this shape for future study and focus on an algebraic one,
\be
\Sigma(u) = (n+1)Au^n
\ee
from which the reflection amplitude calculated
from Eq.~(\ref{eq:reflection}) reads,
\be
\label{eq:mono-CAP}
    r_\Sigma = e^{2ikL}e^{-Ak/E} +
    \frac{An(n+1)(n-1)}{2^{n+2}Ek^nL^{n+1}}
\ee
As a consistency check of the approach developed above, we compare this
analytical result for the reflection probability with direct numerical
calculation using the \textsc{kwant} d.c transport
package~\cite{kwant}. To do so we discretize the continuous Schrödinger
equation onto a lattice of lattice spacing 1.
Figure~\ref{fig:scaling} shows how $R_\Sigma$ scales for the case $n=2$
and $n=6$, showing an excellent agreement between the direct numerical simulations
and the above analytical result in the limit of validity of the latter (small reflection).
Figure~\ref{fig:scaling}c shows that the reflection has a minimum as a function of $A$ which corresponds to a compromise between the first and last term of Eq. (\ref{eq:mono-CAP}). Once $A$
has been chosen large enough for the first term of Eq. (\ref{eq:mono-CAP}) to be negligeable, one can
always choose $L$ large enough to control the second term. We can already anticipate that the difficulties will come from vanishing energies $E \to 0$ for which the spurious reflection goes toward unity.

\subsection{Numerical precision in the time domain}
Now that we understand the d.c. case,
let us consider the previous one dimensional model in the time domain and
send a Gaussian voltage pulse through the wire. This problem has been studied in detailed in
Ref.~\onlinecite{Twave_formalism} to which we refer for more details. We compute the current flowing and measure the error with respect to a reference calculation $I_E^{exact}(t)$,
\be
    \delta = \frac{\int_0^{t_{max}} |I_E(t) - I_E^{ex}(t)|\, dt}{\int_0^T |I_E^{ex}(t)|\, dt}
\ee
where $I_E(t)$ is the time-dependent probability
current for a particle injected at energy $E$ using the above designed imaginary potential to absorb the outgoing waves. The reference calculation is performed without imaginary potential, but with enough
added unit cells in the leads such that the solution does not have time to propagate back into
the central region before the end of the simulation; this corresponds to the WF-C method of
Ref.~\onlinecite{Twave_formalism}.

\begin{figure*}
    \includegraphics[width=\textwidth]{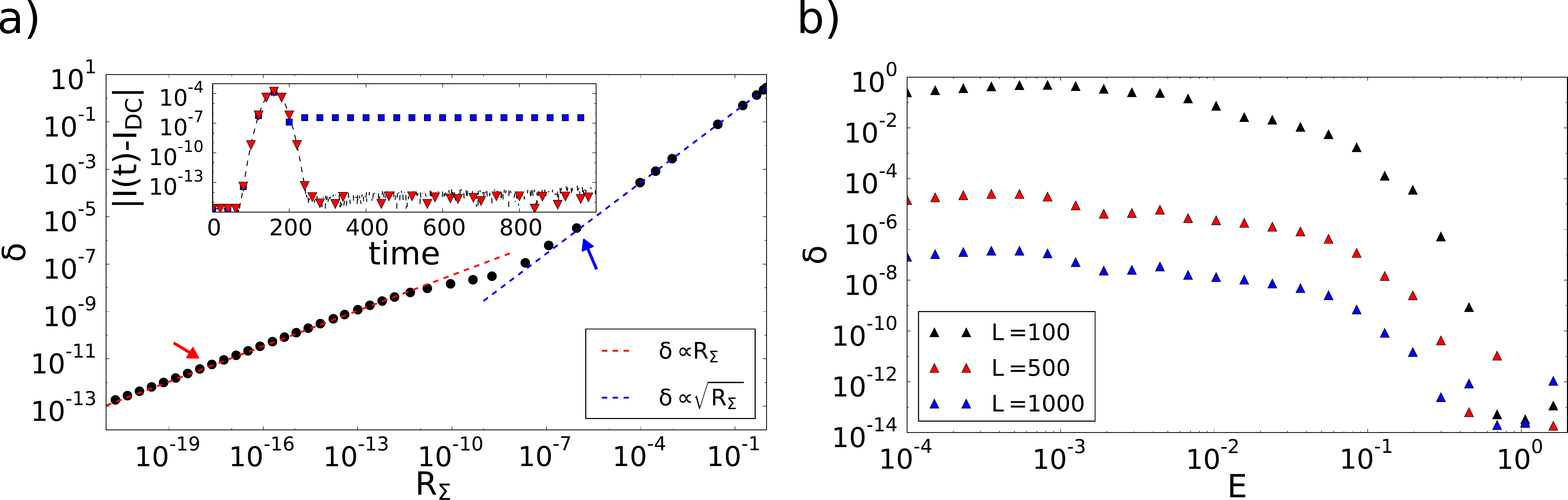}
    \caption{Scaling of the error, $\delta$, in the time-dependent simulation with respect
        to a) the d.c. reflection probability, $R_\Sigma$, and b) the particle injection energy, E.
        A monomial CAP with $n=6$ was used. For a) simulations
        were carried out at a single energy at the centre of the band and the
        length of the absorbing region was varied. Dashed lines show fits to
        $\delta \propto R_\Sigma$ (blue) and $\delta \propto \sqrt{R_\Sigma}$
        (red). \emph{Inset}
        Deviation of the probability current from equilibrium for different
        lengths of the absorbing region corresponding to the two points
        indicated by arrows in the main figure. The black dashed curve shows
        the exact result.
    }
    \label{fig:td-scaling}
\end{figure*}

Figure~\ref{fig:td-scaling}a shows the scaling of the error $\delta$ in the
time-dependent calculation with respect to
the d.c. reflection probability of the absorbing region $R_\Sigma$ as $L$ is
changed. The current at an
energy at the centre of the spectrum is calculated.
  We see from figure~\ref{fig:td-scaling} that for very short
absorbing regions the error scales proportionally to $R_\Sigma$, whereas for longer
regions it scales as $\sqrt{R_\Sigma}$.
This simply reflects the fact that the error on $\bar \Psi_{\alpha E}(t)$ is proportional
to $\sqrt{R_\Sigma} = r_\Sigma$: since the current (hence $\delta$) is quadratic in ${\Psi}_{\alpha E}(t) \equiv e^{-iEt}(\Psi_{\alpha E}^{st} + \bar\Psi_{\alpha E}(t))$, the error has the form $\delta \sim  2|\Psi_{\alpha E}^{st}| \sqrt{R_\Sigma} +R_\Sigma$. More importantly, we see that we can control the error of the calculation with arbitrary precision and for extremely long times (we checked this last point for much longer times than what is shown in the inset).

More interesting is the behavior of the error $\delta$ as a function of the injection energy $E$. Indeed,
since there are large spurious reflections when $E\to 0$, we might expect $\delta$ to behave badly
as one decreases the energy. Figure~\ref{fig:td-scaling}b indeed shows that the error increases as
the energy is lowered. However, one finds that $\delta$ {\it saturates} at small energy. Furthermore,
the saturated valued decreases with $L$ and can thus be controlled. This behaviour comes
from the structure of the wave function as shown in Eq.(\ref{eq:leads}); even though one injects
an electron at a definite energy inside the system, the energy of the outgoing wave is ill defined.
The contribution to the wavefunction coming from spurious reflections takes the form
\begin{dmath}
\label{eq:wavefunc}
\delta\bar\Psi_{\alpha, E}(n, t) =  \int_0^\infty e^{-i(k'n + E't)}\xi_{\alpha'}(E')r_\Sigma(E')\,S_{\alpha' \alpha}(E', E)\, dE'
\end{dmath}
The contribution spreads over an energy window $E_{pulse}$ which characterizes the inelastic scattering matrix,
$S_{\alpha' \alpha}(E',E)$. $S_{\alpha' \alpha}(E',E)$ typically decays on an energy scale of the order of $E_{pulse} = \hbar/\tau_{pulse}$ (see Fig.~10 of Ref.~\onlinecite{Twave_formalism} for an example). For the voltage pulse considered here (which sends one electron through the system), $\tau_{pulse}$ is essentially the duration of the pulse. The consequence is that the reflection $r_\Sigma$ is averaged over an energy window of width $E_{pulse}$, which blurs the $E = 0$ behaviour of $r_\Sigma$:
\be
\delta \approx \langle r_\Sigma (E) \rangle_{E<E_{pulse}} \approx r_\Sigma (E_{pulse})
\ee
We conclude that the error can always be made arbitrarily small,
irrespective of the duration of the simulation. A slight drawback is that for a given
imaginary potential, the precision of the calculation can depend on the actual physics taking place inside the central system (which sets $E_{pulse}$) if one injects electrons with energies close to the band edges of the leads.

\subsection{A general algorithm}
We now discuss how to turn the above results into a practical scheme to perform numerical calculations in a robust way.

Since we cannot guarantee the error for a given shape of the imaginary potential (we have seen above that it might depend on the physics of the central region), we first need to design an algorithm for an on-fly calculation of an error estimate (without the reference calculation used above). This can be done
as follow for a small additional computational cost. In the integration of the Schrödinger equation,
one separates the wave function in the central region $\bar\psi_{\bar 0}$ and in the leads
$\bar\psi_{\bar 1}$ (let us suppose that there is only one lead for simplicity). The equations to be integrated take the block form,
\begin{eqnarray}
i\partial_t \bar\psi_{\bar 0} &=& H_{\bar 0\bar 0}(t) \bar\psi_{\bar 0} + H_{\bar 0\bar 1} \bar\psi_{\bar 1} + S_{\bar 0}(t) \\
i\partial_t \bar\psi_{\bar 1} &=& H_{\bar 1\bar 1}(\Sigma) \bar\psi_{\bar 1} + H_{\bar 1\bar 0} \bar\psi_{\bar 0}
\end{eqnarray}
where $S_{\bar 0}(t)$ is the source term present in the central region
and the imaginary potential is included in $H_{\bar 1\bar 1}$. One then introduces a second ``copy'' of
the lead wave function $\bar\psi_{\bar 1}'$ that uses a different imaginary potential
$H_{\bar 1\bar 1}(\Sigma')$. The equations of motion for this ``copy'' are
\begin{eqnarray}
i\partial_t \bar\psi_{\bar 1}' &=& H_{\bar 1\bar 1}(\Sigma') \bar\psi_{\bar 1}' + H_{\bar 1\bar 0} \bar\psi_{\bar 0}
\end{eqnarray}
One then keeps track of both $\bar \psi_{\bar1}$ and $\bar \psi'_{\bar1}$ simultaneously, although only
$\bar \psi_{\bar 1}$ will affect the dynamics of $\bar \psi_{\bar 0}$.
The trick is to design $\Sigma'(n)=\Sigma(n-M)$, i.e. to insert $M$ extra lead layers before the
imaginary potential, and to monitor the difference between $\bar\psi'_{\bar 1}$ and $\bar\psi_{\bar 1}$
in the lead cell adjacent to the central region,
$\delta\bar\psi_{\bar 1} = \bar\psi_{\bar 1} - \bar\psi'_{\bar 1}$. Spurious reflections from
the presence of $\Sigma$ will arrive at the boundary of the central region for $\bar\psi_{\bar 1}$
\emph{before} $\bar\psi'_{\bar 1}$, as the latter has $M$ extra lead layers. This delay in the arrival
of the spurious reflections will give rise to a finite $\delta\bar\psi_{\bar 1}$. Note that
$\delta\bar\psi_{\bar 1}$ will remain 0 in the case that there are no spurious reflections.
$\delta\bar\psi_{\bar 1}$ can thus be used as an error estimate for the wavefunction in the lead.

In the worst case scenario this scheme will increase the computational cost by a factor of 2 (when
the absorbing region represents the largest part of the system). It is worth noting, however, that
without an error estimate for the spurious reflections one would have to check for convergence of
results by performing several simulations with different values of $L$, the absorbing region length.

The remaining task is to choose the parameters $A$ and $L$ for a given shape of the imaginary potential. Ideally we would choose $L$ as small as possible so as to minimize the extra computational effort while requiring that $|\delta\bar\psi_{\bar 1}|$ remain smaller than a fixed maximum error, $\delta_{max}$.
Given  $\delta_{max}$ it is easy to choose $A$ such that the first term in Eq.~\eqref{eq:mono-CAP}
is not a limitation. By noting that $e^{-Ak/E} < e^{-A/(aB)}$ ($B$ is the lead bandwidth and $a$ is the
discretization step) we see that it is sufficient to choose $A$ such that $e^{-A/(aB)}<\delta_{max}$ for the absorption process not to be the limiting factor of the precision.
Next, one needs to choose $L$ large enough to enforce
$|\delta\bar\psi_{\bar 1}| < \delta_{max}$. In practice, we found that a few hundred (up to a thousand) lead cells
is almost always sufficient for the physics we have studied so far, for typical $\delta_{max} \sim 10^{-5}$.

Let us end with a last point of practical importance. We have seen that the
major contribution to spurious reflection comes from a narrow region around the
band edge of the lead. The wavefunctions associated with these energies
propagate \emph{extremely} slowly into the absorbing region due to the
vanishing velocity at the band edge. Unless one is interested in extremely long
simulation times, we can take advantage of this by placing a small number of
lead layers before the imaginary potential. The slow-moving waves will induce
spurious reflections, but will take a long time to traverse this buffer layer
due to their small group velocity. Meanwhile the absorbing region does not have
to be made as long, as it does not have to absorb waves of vanishingly small energy.

\section{Voltage pulses in long Josephson Junctions}
\label{sec:jj}

We are now in possession of a robust algorithm to simulate time-dependent open systems.
Compared with our own previous approach, the computing time is now $\order{N t_{max}}$.
This new algorithm allows us to treat cases where very
small energies (hence large times) come into play. We now turn to a specific application
concerning superconducting - normal - superconducting Josephson junctions where the large
separation of scales $E_{th} \ll \Delta_0 \ll E_F$ ($E_{th}$: Thouless energy, $\Delta$: superconducting gap, $E_F$: Fermi energy) makes a linear scaling algorithm very welcome.
An interesting aspect of superconductivity is that the problem is intrinsically time dependent
{\it even in d.c.} as soon as there are voltage differences across the superconductors
(as evidenced by the a.c. Josephson effect which transforms a d.c. voltage into an a.c. current).
We emphasize that the algorithm is in no way limited to superconductivity and refer to
the introduction for other applications (such as quantum Hall effect, graphene...)

In the following, we will focus on 3 physical effects. First we will recover known physics of Josephson junctions: the Multiple Andreev Reflection (MAR) phenomena and the a.c. Josephson effect.
Second, we will discuss the relaxation of a SNS junction after an abrupt raise of the applied potential, showing how MAR comes into play in the relaxation rate. Third, we will study a novel
phenomenon, the propagation of a voltage pulse through a Josephson junction.

\subsection{Minimum microscopic model for a SNS junction}
We consider voltage-biased Josephson junctions.
In this setup we have two infinite superconducting reservoirs coupled by a normal region
of length $L$.  We shall treat the problem using a 1D Bogoliubov-De-Gennes Hamiltonian\cite{deGennes}:
\begin{equation}
    \label{eq:jj_hamiltonian}
    \hat H = \int_{-\infty}^{\infty} \hat{\mathbf{\Psi}}^\dagger(x)
        \left(\begin{array}{cc}
            \frac{p^2}{2m} - \mu(x,t) & \Delta(x, t) \\
                \Delta(x, t)^* & \mu(x,t) - \frac{p^2}{2m} \\
        \end{array}\right)
        \hat{\mathbf{\Psi}}(x)\;
        dx
\end{equation}
where $p = -i\hbar\frac{\partial}{\partial x}$,
$\hat{\mathbf{\Psi}}(x) = (\hat{\psi}_\uparrow(x),\, \hat{\psi}_\downarrow^\dagger(x))^T$ and $\hat{\psi}_\uparrow(x)$
is an operator which annihilates an electron at position $x$ in a spin up state.
$\Delta(x, t)$ is the superconducting order parameter, which reads,
\be
 \Delta(x, t) = \left\{
 \begin{array}{l}
 \Delta_0 \ {\rm for }\  x > L \\
 0 \ {\rm for }\  0 \le x \le L \\
 \Delta_0 \exp[-2i\phi(t)]\ {\rm for }\  x < 0
 \end{array}
 \right.
\ee
with $\phi(t) = (e/\hbar)\int_0^t V_b(\tau)\, d\tau$ and $V_b(t)$ is
the voltage bias applied to the left superconductor (which is 0 before $t=0$).
Likewise, the electrical-chemical potential $\mu(x,t)$ reads,
\be
 \mu(x, t) = \left\{
 \begin{array}{l}
 E_F + V_b(t) \ {\rm for }\  x \le 0 \\
 E_F + U(x)  \ {\rm for }\  0 \le x \le L \\
 E_F \ {\rm for }\  x > L
 \end{array}
 \right.
\ee
where $E_F$ is the Fermi energy and $U(x)$ a potential barrier. We only consider a single spin sector as our model is spin independent; the two spin
sectors give degenerate solutions.

In order to put eq.~(\ref{eq:jj_hamiltonian}) into a form where we can apply the algorithm developed above we first apply a gauge transformation
\begin{equation}
    \hat{\mathbf{\Psi}}'(x) = (\Theta(x) + \Theta(-x)\exp[i\phi(t)\boldsymbol{\tau}_z])\hat{\mathbf{\Psi}}(x)
\end{equation}
where $\boldsymbol{\tau}_{\{x,y,z\}}$ are Pauli matrices and $\Theta(x)$ is the Heaviside function. This transformation brings all the time-dependence for $x < 0$ into a time-dependence in the momentum term at $x=0$, the boundary between the left superconductor and the normal region. In this gauge
both superconductors are at equilibrium. We next discretize onto a lattice with spacing $a$,
using a central difference approximation for the second spatial derivative,
$\partial^2 \Psi / \partial y^2 \approx [\Psi(y + a) + \Psi(y - a) - 2\Psi(y)]/a^2$,
to obtain a tight-binding model:
\begin{equation}
    \label{eq:discretized_jj_hamiltonian}
    \hat H_{tb} = \sum_{i,j=-\infty}^{\infty} \hat{\mathbf{c}}^\dagger_i \mathbf{H}_{i,j}(t) \hat{\mathbf{c}}_j
\end{equation}
with the matrices $\mathbf{H}_{i,j}(t)$ being non-zero only for diagonal and nearest-neighbour matrix elements,
\begin{equation}
    \mathbf{H}_{j,j}(t) = \left[\frac{\hbar^2}{ma^2} - E_F +U_j \right]\boldsymbol{\tau}_z
                          + \Delta_0(\theta_{0, j} + \theta_{j, L})\boldsymbol{\tau}_x
\end{equation}
\begin{equation}
    \mathbf{H}_{j,j+1}(t) = \frac{-\hbar^2}{2ma^2}\boldsymbol{\tau}_z
                            \exp\left[i\phi(t)\delta_{j, 0}\boldsymbol{\tau}_z\right]
\end{equation}
\begin{equation}
    \mathbf{H}_{j,j-1}(t) = [\mathbf{H}_{j, j+1}(t)]^\dagger
\end{equation}

where $\hat{\mathbf{c}}_j \equiv \hat{\mathbf{\Psi}}(ja) = (\hat{\psi}_\uparrow(ja), \hat{\psi}^\dagger_\downarrow(ja))^T$
(and $\hat{\mathbf{c}}^\dagger_j$, its Hermitian conjugate) are vectors of creation (annihilation) operators at site $j$.
$\delta_{i, j}$ is the Kronecker delta
and $\theta_{i, j}$ is a discrete Heaviside function, defined as 1 if $i>j$ and 0 otherwise. $U_i$ is the potential barrier. This model can be readily solved numerically using the above-developed technique.

\subsection{Multiple Andreev Reflection and a.c. Josephson Effect}
\begin{figure}
    \includegraphics[width=0.48\textwidth]{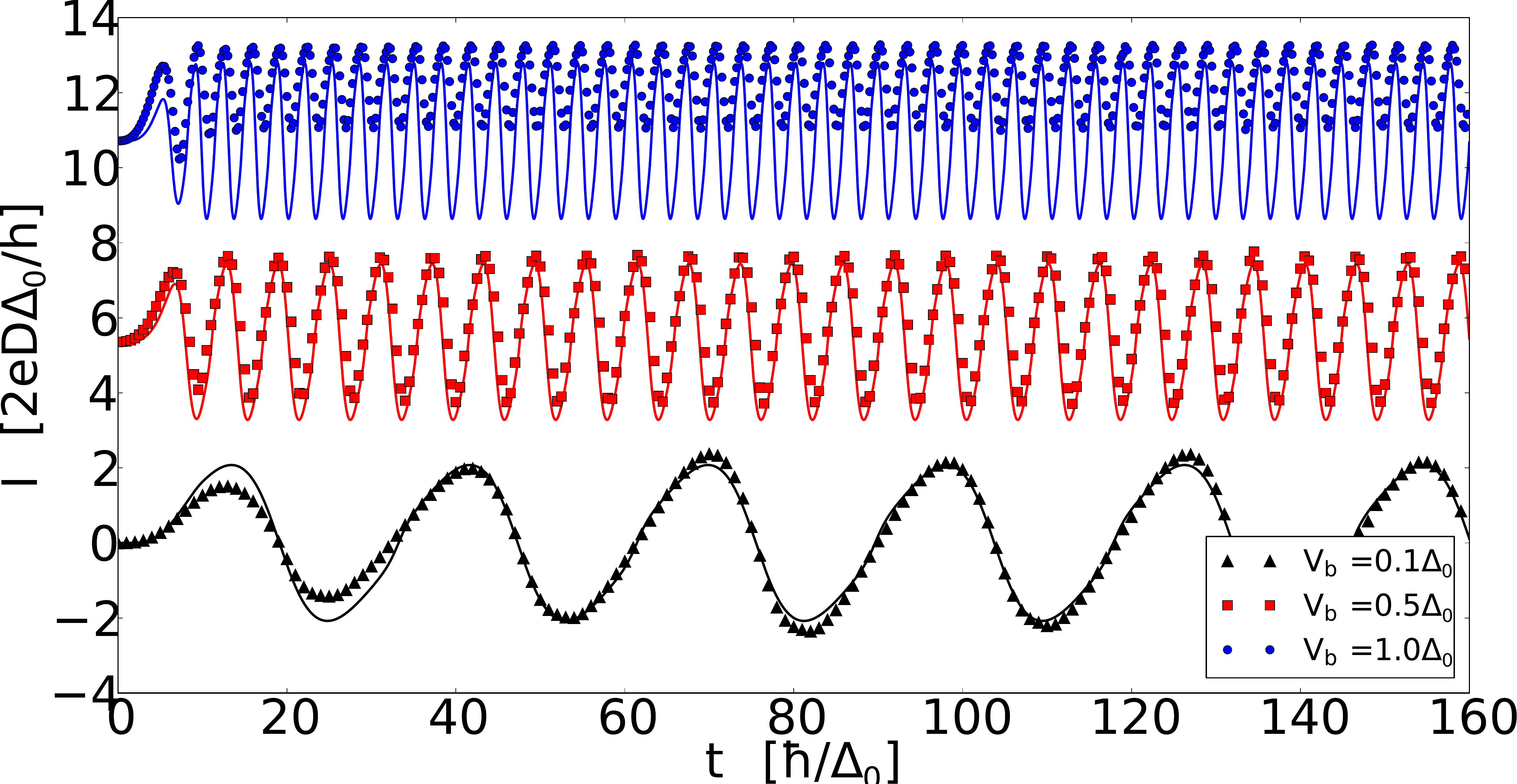}
    \caption{The a.c. Josephson effect. The different curves show the calculated current as a function
        of time for different bias voltages across a short junction with a transmission of 0.7. The full curves and symbols show the theoretical and
        numerical results respectively. The curves have been vertically offset for clarity.
    }
    \label{fig:AC_jos}
\end{figure}
Let us now apply our numerical technique and discuss the physics of a voltage biased Josephson junction.
There are two very different regimes to discuss: at low voltage one observes the a.c. Josephson effect,
while at higher voltage one observes multiple Andreev reflections (MAR). Both effects
are closely related, as the Josephson effect corresponds to the limit of an infinite number of Andreev
reflections, yet they are usually calculated with different techniques. Indeed, one of the
challenges of such a simulation is that to access small bias voltages $V_b$ one needs to go to very long times $\propto \hbar/V_b$. For this problem the source-sink algorithm thus has a distinct
advantage over previous methods due to its linear scaling with simulation time.

In this subsection, we concentrate on
a short junction and add a potential barrier $U(x)$ that allows us to tune the transmission probability $D$
of the normal part of the junction from insulating $D\ll 1$ to ballistic $D=1$.
To obtain a current-voltage characteristic for the junction we perform a separate
simulation for each value of voltage required. For a given simulation (voltage value) we use the
following protocol. At $t=0$ the voltage of the left superconductor is raised smoothly,
$V_b(t) = (V_0/2)(1 - \cos(\pi t/T))$, until $t=T$, when $V_b$ is held at a value $V_0$ (we used
$T = 50\,\hbar/\Delta$). The system relaxes to a steady state and we can obtain the current
using eq.~(\ref{eq:time-dep-obs}). The d.c. current can then be obtained by taking an average over one period
of the fully time-dependent current after the system has reached a steady state.

Let us start with the a.c. Josephson effect. At equilibrium, the ground state energy $E(\phi)$ of the junction depends on the phase difference $\phi$ between the order parameters of the two superconductors. The corresponding supercurrent is given by
\begin{equation}
\label{eq:adiabatic_superconductor}
I = (2e/\hbar)  \partial E/\partial\phi \propto \cos\phi.
\end{equation}
When a small bias is applied to the junction, $\phi$ increases linearly in time $\phi(t)=2e V_bt/\hbar$ and one observes the a.c. Josephson effect
at frequency $2eV_b/h$. This is perhaps the most striking manifestation of superconductivity; a d.c. bias leads to an a.c. effect. Figure \ref{fig:AC_jos} shows a numerical calculation of  the current as a function of time together with the adiabatic prediction discussed above (the dispersion relation $E(\phi)$ was calculated from the equilibrium junction and differentiated numerically). We see a perfect agreement at low bias, indicating that our technique can reach the adiabatic limit. Upon increasing the bias, one leaves the adiabatic limit and the corresponding prediction becomes less accurate.

\begin{figure}
    \includegraphics[width=0.48\textwidth]{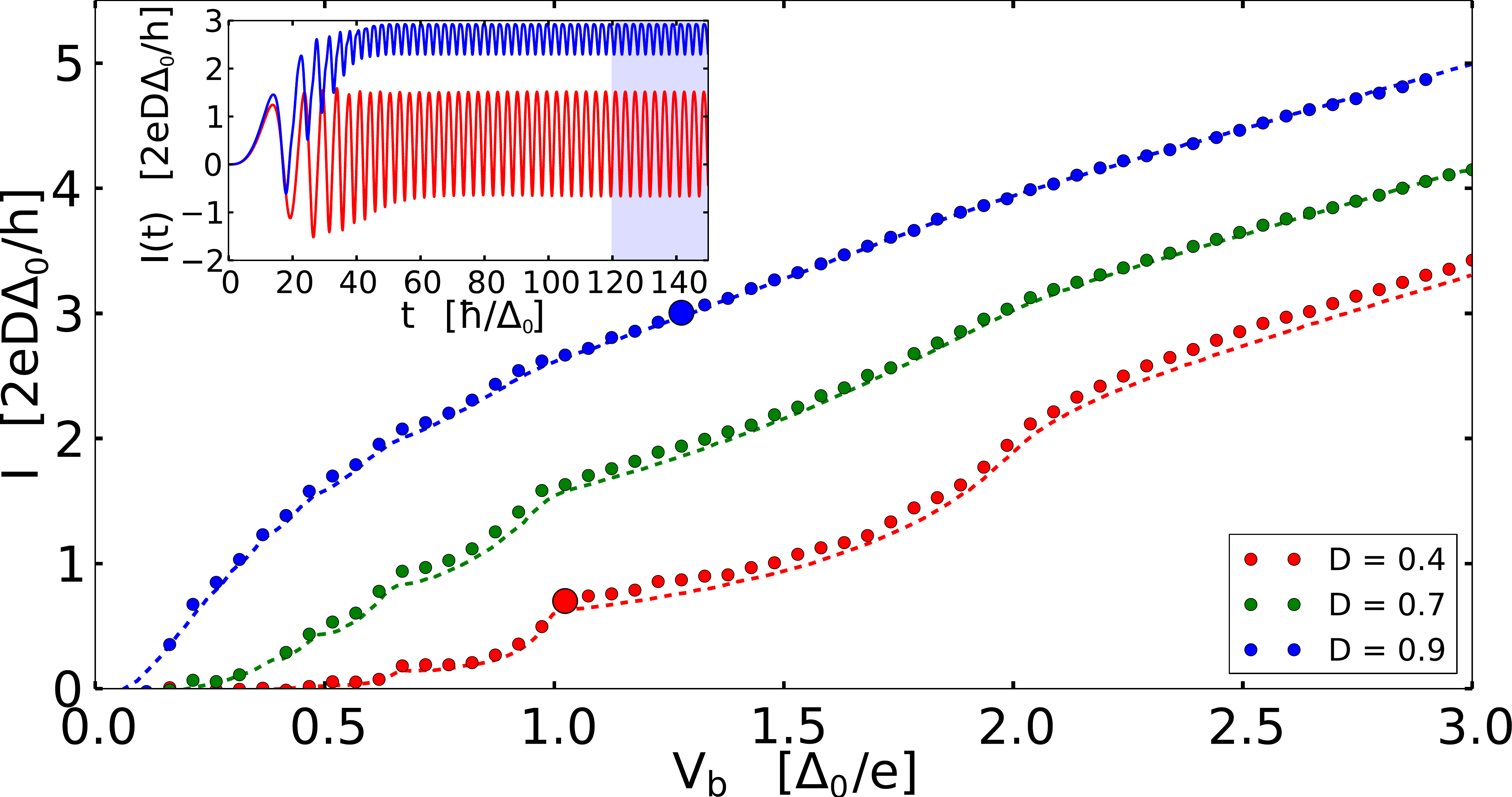}
    \caption{d.c. current-voltage curve showing the analytical results from
        Ref.~\onlinecite{avarin_MAR} (dashed line) and the source-sink
        numerical calculation (points) for different values of the transmission
        ($D$) of the insulating link. Inset: time series corresponding to the
        enlarged points in the main figure, showing a typical averaging window
        over which the d.c. current was calculated.
    }
    \label{fig:MAR}
\end{figure}

Indeed, as one increases the bias, a d.c. component starts to appear in the current.
This is best understood starting from large bias. For $V_b> 2\Delta_0/e$, the charges can flow directly from the left ``valence'' band of the superconductor to the right ``conduction'' band (using the semiconductor terminology). As one lowers the bias, this
direct process is no longer possible and at least one Andreev reflection takes place on the right superconductor. As
one further lowers the bias, more and more Andreev reflections are needed and one observes kink in the I-V characteristics at
values $V_b> 2\Delta_0/Ne$ with $N=1,2,3...$ The Fourier components of the MAR current have been previously calculated using a Floquet approach\cite{avarin_MAR, cuevas_MAR} and are routinely observed experimentally (see for instance Ref.~\onlinecite{MAR_break_junction}). Here we recover those results using a microscopic model
for the junction. Figure~\ref{fig:MAR} compares the current-voltage characteristics of such a
junction calculated in Ref.~\onlinecite{avarin_MAR} with a
simulation using the source-sink algorithm for different values of the
transmission ($D$) of the junction. We see a very good agreement with these previous results.

\begin{figure}
    \includegraphics[width=0.48\textwidth]{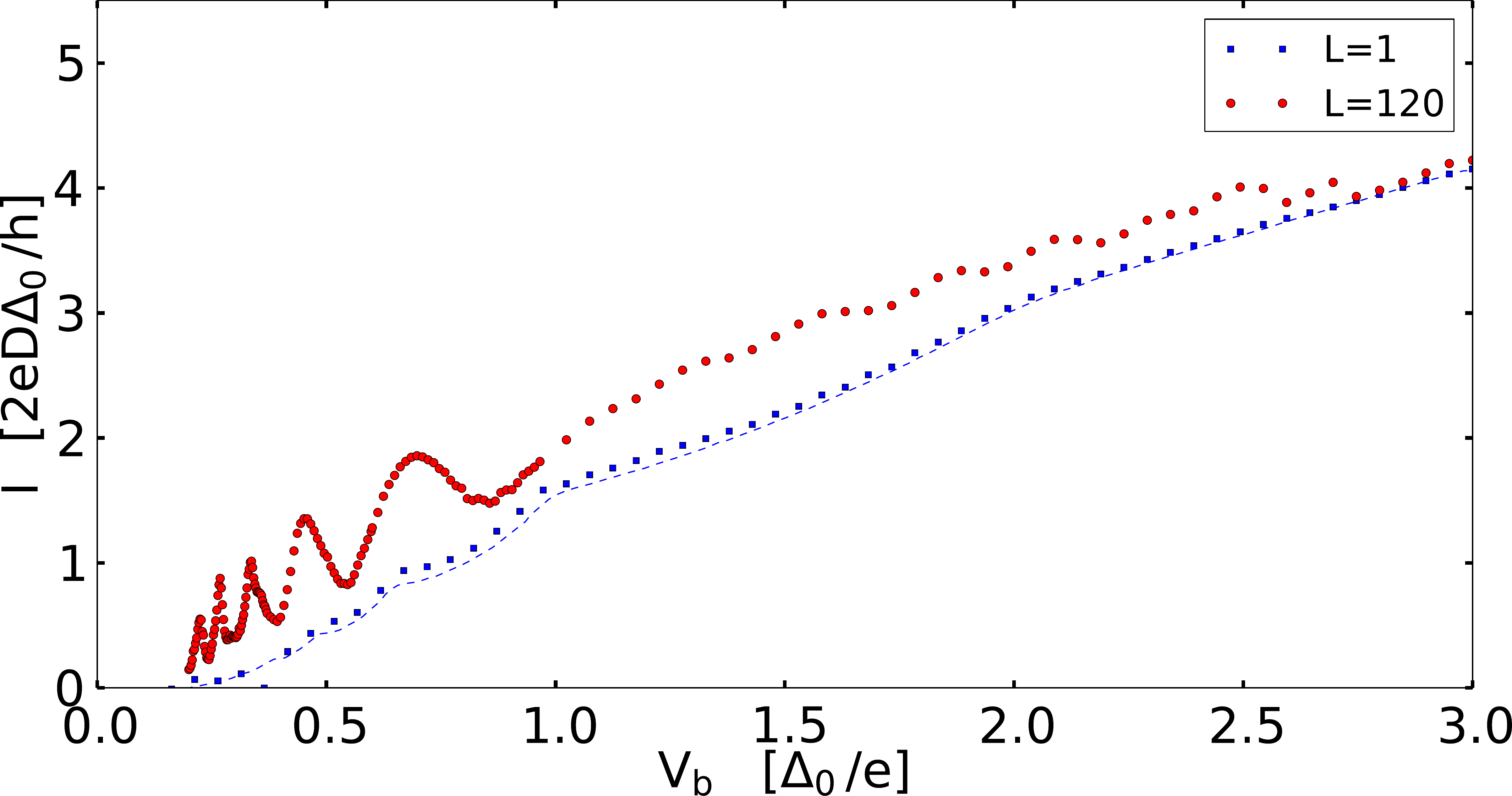}
    \caption{
        Comparison of the current-voltage characteristics for a short junction (one site
        in the normal region) and a long junction (120 sites in the normal region).
        Both the junctions have a transmission of 0.7.
    }
    \label{fig:long_MAR}
\end{figure}

Using the source-sink algorithm we can go beyond the limitations of an analytical approach for
little extra overhead. We can, for example, explore the behaviour of a long Josephson junction
under voltage bias. Figure~\ref{fig:long_MAR} compares the current-voltage characteristics of
a long junction with the short junction studied previously. We clearly see that the long junction
has more sub-gap features, which can be attributed to the larger number of Andreev states
below the gap. We see that numerics has an advantage over analytical approaches in this regard,
in that it is relatively cheap to explore new regions of parameter space or in crossover regions between
tractable limits (e.g. short junction vs. long junction).

\subsection{Relaxation of Andreev bound states}
An important difference of Josephson junctions with respect to
other nanoelectronics systems is the presence of (so called Andreev) bound states.
Since Andreev states have their energies inside the superconducting gap, there
is no continuum band with which they can hybridize so that they have infinite lifetime.
These states must be added explicitly in Eq.~\eqref{eq:time-dep-obs} and the
definitions of the Green's function (see section IV of Ref.~\onlinecite{profumo_quantum_2015} for
a discussion). As the Andreev states carry the Josephson current, their role is
particularly important and they cannot be ignored. This is in contrast to many non-superconducting
systems where the bound states do not contribute to transport.

Andreev states give us another opportunity to study MAR physics. Suppose that we abruptly
raise the voltage bias at $t=0$, thereby placing the system in a non equilibrium state.
Just after the voltage raise, a given wavefunction can be decomposed on the eigenbasis of the
equilibrium SNS junction,
\be
\Psi = \int dE \ c(E)\  \Psi_{\alpha E}^{st}  + \sum_n c_n \Psi_{n}^{st}
\ee
where $c(E)$ and $c_n$ are respectively the projection of the wave function on the scattering states
and the bound states ($\Psi_{n}^{st}$). It is important to realize that in the absence of bias voltage, the bound state part of the wave function will {\it never} relax (within the above model)
as the Andreev states are true bound states with energy $E_n$: the second part of the wave function will simply oscillate as $\sum_n c_n e^{-iE_nt} \Psi_{n}^{st}$ for ever.
However, the presence of the bias voltage allows the energy to change by $eV$ in between two Andreev reflections so that after $N\approx \Delta_0/(eV_b)$ reflections, one can reach energies outside the gap
and the wavefunction can relax. Denoting $\tau_P = L/v_F$ the time of flight between two Andreev reflections, we expect the relaxation time $\tau_R$ of the system to behave as $\tau_R \propto
N \tau_F = L\Delta_0/(v_F eV_b)$.

\begin{figure}
    \includegraphics[width=0.48\textwidth]{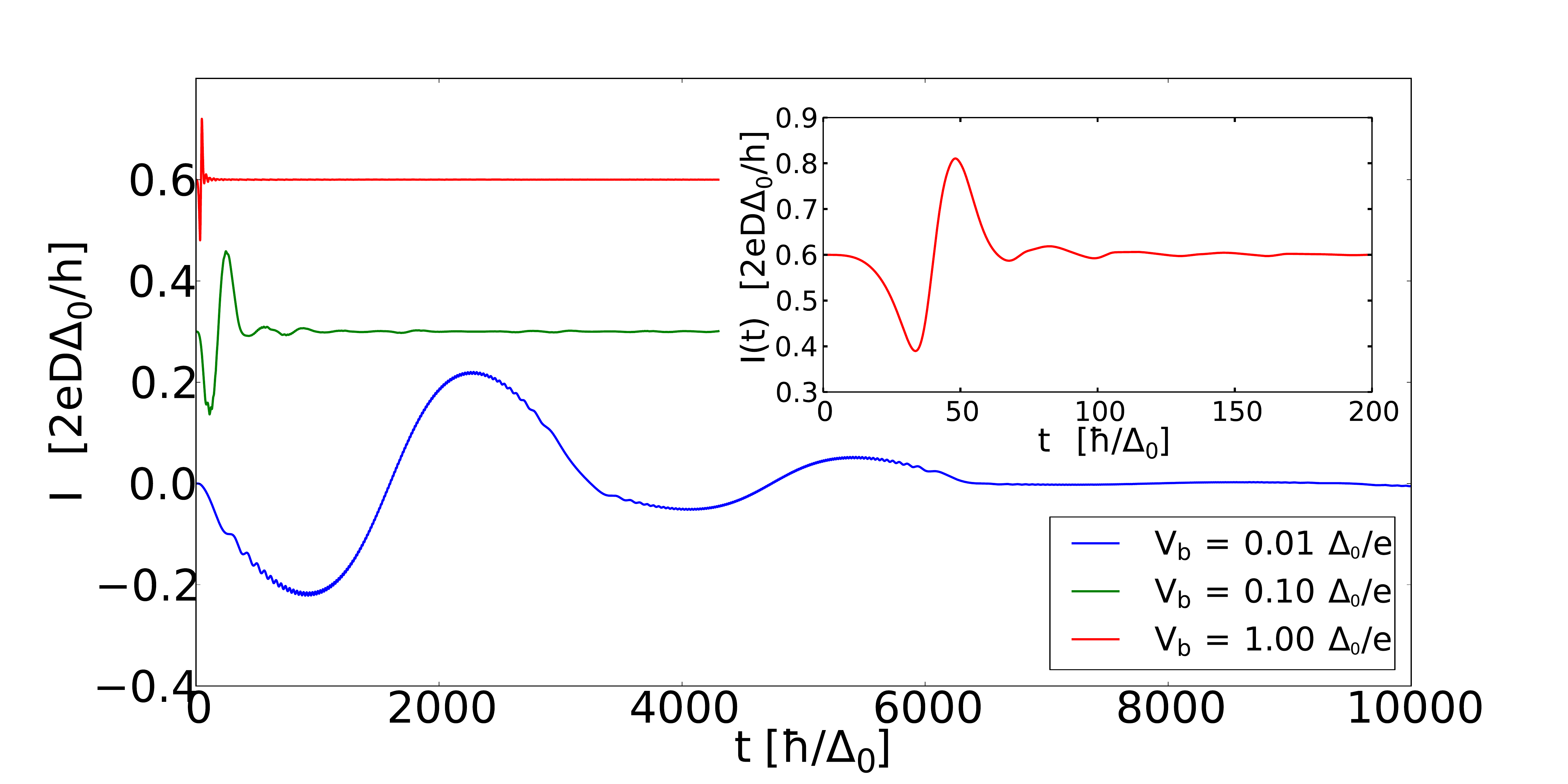}
    \caption{
        Current contribution from the (Andreev) bound states at different bias voltages.
        The curves have been offset for clarity. The inset shows a zoom of the curve
        for $V_b = \Delta/e$.
    }
    \label{fig:MAR_boundstate}
\end{figure}

Figure~\ref{fig:MAR_boundstate} shows the contribution of the
Andreev bound states to the current as a function of time for three values of the bias voltage.
We indeed see that the current carried by the bound states dies away with time in presence
of a finite bias. Although we did not try to define $\tau_R$ precisely, we clearly see that dividing $V_b$ by a factor 10 leads to a 10 time increase of the relaxation time, establishing the relation $\tau_R \propto 1/V_b$ which originates from the MAR assisted relaxation process.

From a numerical perspective, we note that these simulations are taken to extremely long
times, $10^5$ in units of the inverse hopping parameter, $\gamma\,(= \hbar^2/2m^*a^2)$, of the model
(we chose $\Delta=0.1\gamma$ for the above calculations). This calculation clearly necessitates the
source-sink algorithm; we used an imaginary absorbing potential of order $n=6$ with 1000 lead cells forming the
absorbing layer.

\subsection{Propagation of a voltage pulse through a Josephson junction}
A natural consequence of the above discussion is that if one sends a fast voltage
\emph{pulse} through the system (i.e. the final bias voltage vanishes instead of having a finite value),
then the corresponding bound state contribution will not relax and will oscillate for ever
(within the assumptions of our model).

Let us study the corresponding protocol. We consider a perfectly transparent junction with a finite
width, and apply a Gaussian voltage pulse of duration $\tau_P$ on the left
superconducting contact. The junction has a length $L$ such that the time of flight
is $\tau_F = L / v_F$. We consider a ``long'' junction, such that $\Delta_0 \tau_F/\hbar \gg 1$.
We further consider fast pulses with $\tau_F/\tau_P \gg 1$
($\tau_F/\tau_P \sim 5$ in our case). The case of slow pulses is trivial as the physics is essentially given by the adiabatic limit.
The physics of fast pulses is simple yet rather interesting. The pulse generates an electron-like excitation that propagates
through the system until it reaches the right superconductor. There, it is Andreev reflected as a hole-like excitation and a Cooper pair is generated in the right electrode.
The excitation now propagates backward towards the left superconducting electrode where it is Andreev reflected a second time (and a Cooper pair is absorbed from the electrode). The excitation then continues
its propagation again to the right. Within the above model, nothing stops this process and the excitation continues to oscillate back and forth for ever. This is rather appealing: one sends a short voltage pulse and gets an oscillating current at frequency $1/(2\tau_F)$. Beyond the current model, the
relaxation time of the system will be given by the fluctuations of the voltage due to the electromagnetic environment and we anticipate a relaxation of the current on a scale given by the corresponding $RC$ time.

\begin{figure}
    \includegraphics[width=0.48\textwidth]{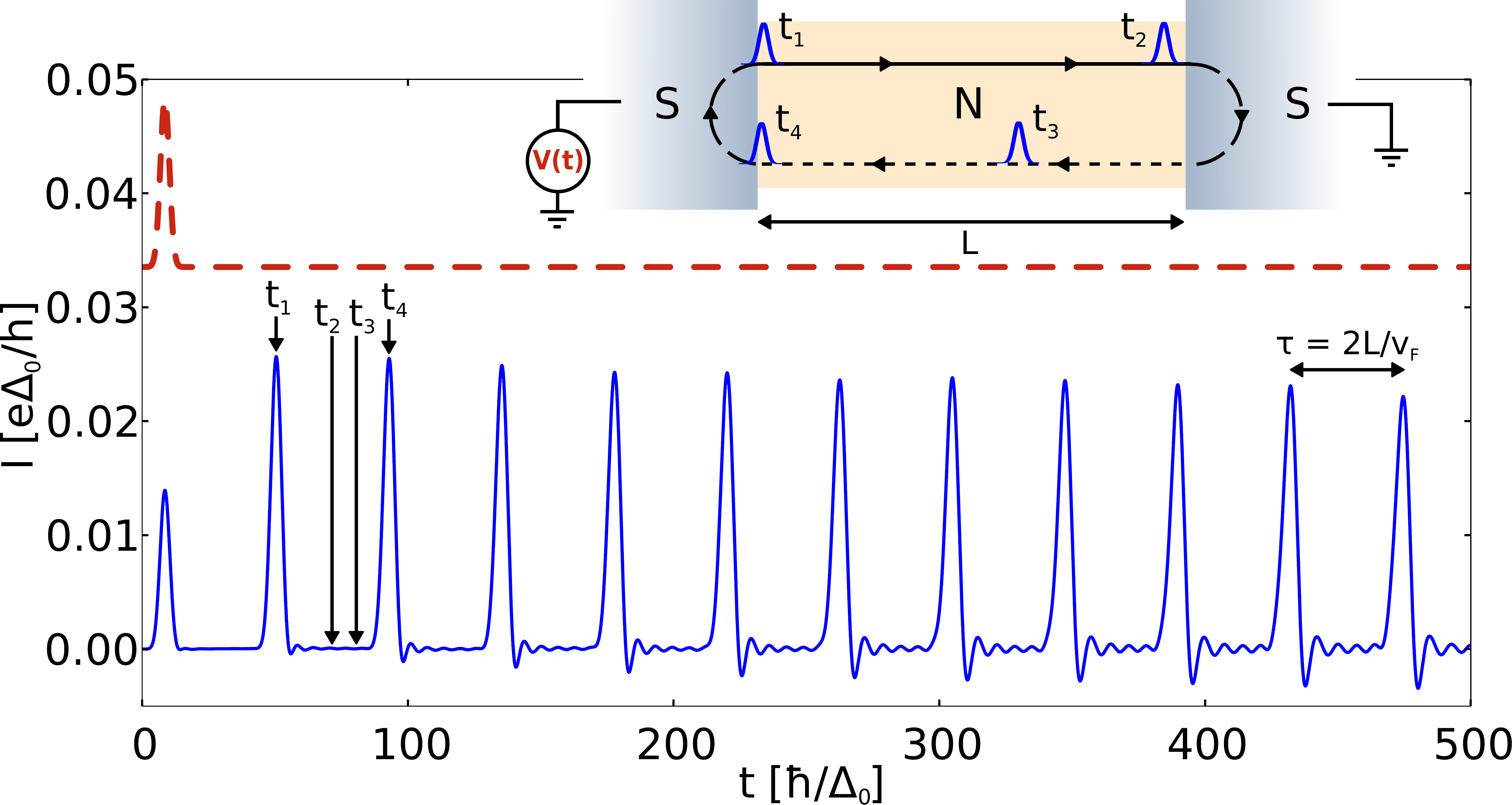}
    \caption{
        Current (blue full line) and voltage (red dashed line, offset for clarity) at the left superconducting-normal contact
        as a function of time. Inset: propagation of the charge pulse through the junction at different times
        ($t_1$, $t_2$, $t_3$, $t_4$) and the corresponding times indicated on the main plot.
    }
    \label{fig:DAR}
\end{figure}

Figure~\ref{fig:DAR} shows a numerical simulation of the propagation of a voltage pulse
as discussed above. Despite the fact that there is only a single voltage pulse at the start, we see pulses of current every $2\tau_F$. We do not observe any quasiparticle current in the
superconducting lead; this (super)current is purely associated with the Andreev reflection process described above.

\begin{figure}
    \includegraphics[width=0.48\textwidth]{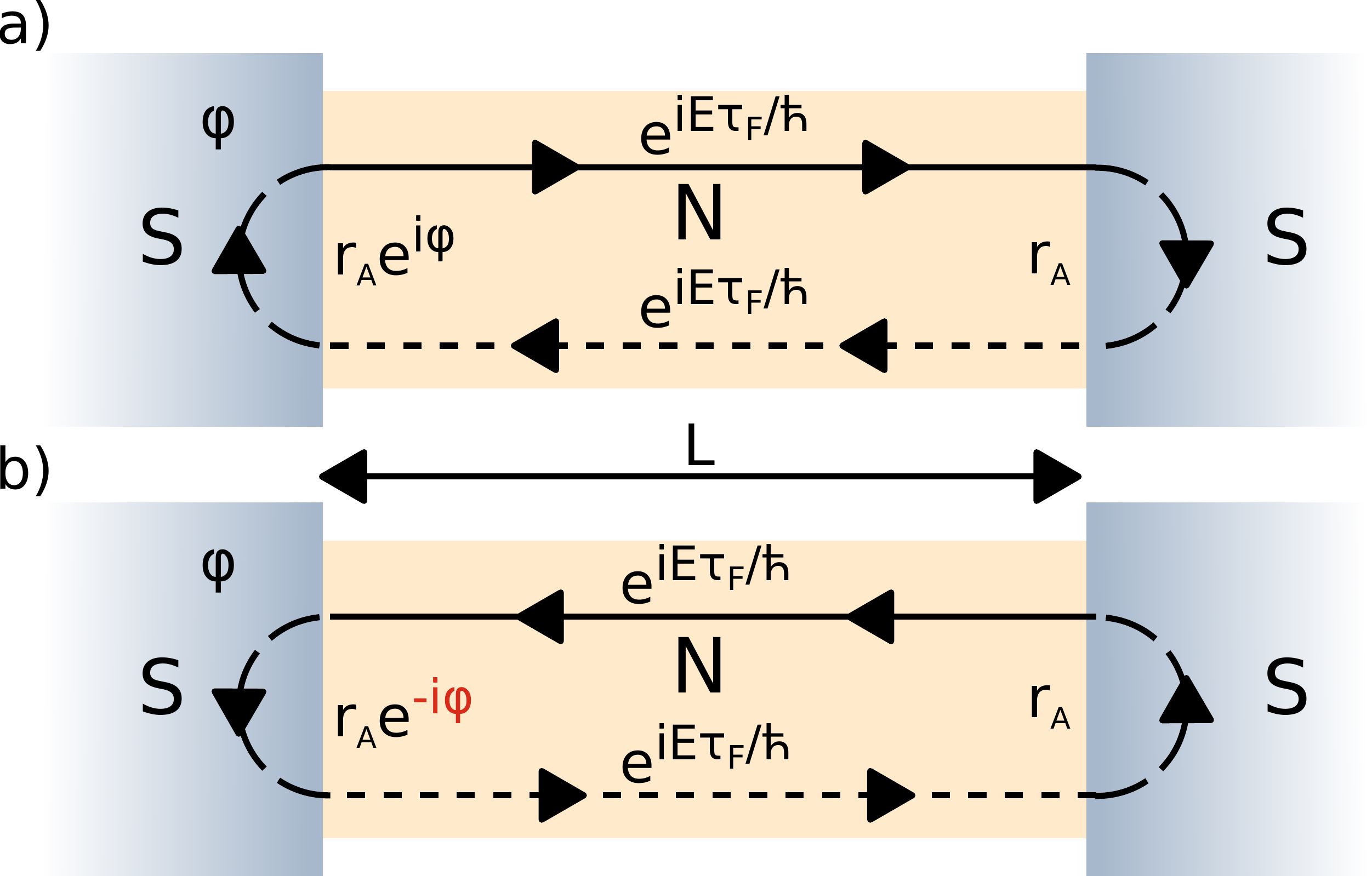}
    \caption{Sketches of the two classes of paths that can result in bound states.
    The full lines corresponds to an electron-like excitation, and the dashed line
    to a hole-like one. Andreev reflection at the normal-superconductor interface
    converts an electron-like excitation to a hole-like one.
    Each sketch actually represents a set of paths with 1, 2, 3, \ldots pairs
    of Andreev reflections.}
    \label{fig:paths}
\end{figure}

\begin{figure}
    \includegraphics[width=0.48\textwidth]{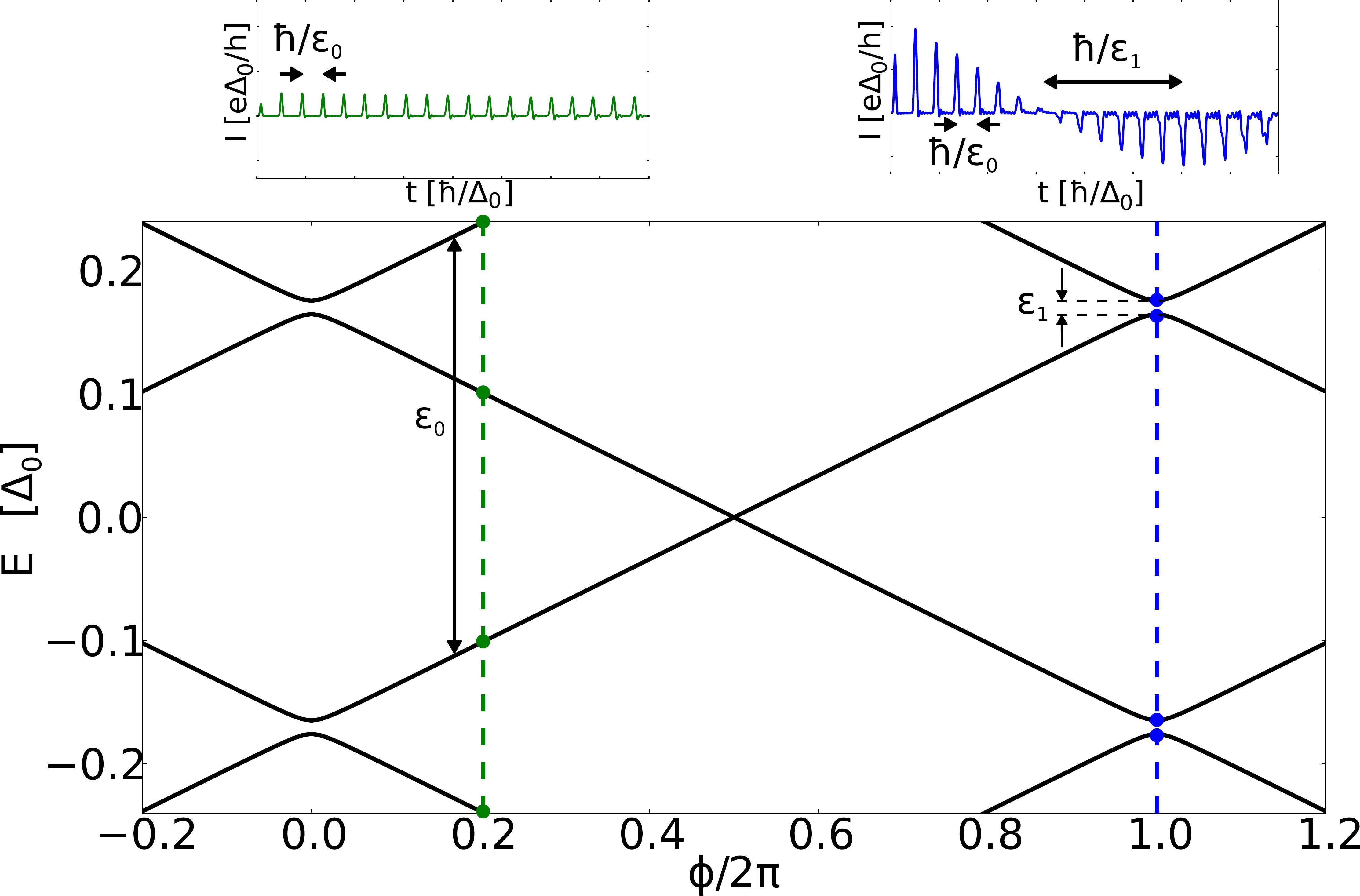}
    \caption{
        A section around $E=0$ of the bound state spectrum after the passage of
        a pulse as a function of the phase $\phi$ picked up from the pulse.
        The vertical dashed lines highlight the bound state energies
        for two values of $\phi$. The current flowing through the junction
        as a function of time is shown in the traces above the main figure.
        The spectrum was calculated numerically by diagonalizing the
        Hamiltonian of the system projected onto a large, finite region around
        the junction
    }
    \label{fig:bound_spectrum}
\end{figure}

We can go a little bit further and look at the structure of the bound states that carry the supercurrent.
They are given by the stationary condition\cite{weston_manipulating_2015, shuo_proposal_2013, beenaker_andreev_1992},
\begin{equation}
r_A^2e^{2iE\tau_F/\hbar}e^{i\phi} = 1
\end{equation}
where left-hand superconductor is at a phase bias $\phi$
compared to the right-hand one and $r_A = E/\Delta_0 - i\sqrt{1 - (E/\Delta_0)^2} $ is the Andreev reflection
amplitude for a particle incident on the superconductor at energy $E$.
The paths contributing to this amplitude are sketched
in Fig.~\ref{fig:paths}a. A similar expression exists for the reversed paths
where the sign of $\phi$ is flipped; this is sketched in Fig.~\ref{fig:paths}b.
For $E < \Delta_0$ we have $r_A = e^{-i\arccos(E/\Delta_0)}$, and we can re-write
this condition as
\begin{equation}
  -2\arccos(E/\Delta_0) + \frac{2E\tau_F}{\hbar} \pm \phi = 2\pi m \ ,\quad m\in\mathbb{Z}.
\end{equation}
In the long junction limit ($\Delta_0 \gg \hbar/\tau_F$) close to zero energy this simplifies to:
\begin{equation}
E = \frac{h}{2\tau_F} \left[m + \frac{1}{2} \mp \frac{\phi}{2\pi}\right]
\end{equation}
which corresponds to two set of equidistant energies separated by $\hbar/(2\tau_F)$, one set
that has energy increasing with $\phi$, and the other decreasing with $\phi$.
Each of these sets corresponds to a ballistic propagation in the continuum limit $\tau_P \ll \tau_F$.
The numerical spectrum, which is shown in Fig.~\ref{fig:bound_spectrum}, adheres to the above-derived result except near the degeneracy points.
The degeneracies are lifted due to the finite ratio $\Delta_0/E_F$ used in the numerical calculation,
which induces a finite normal reflection at the normal-superconducting interfaces.
The two insets of Fig.~\ref{fig:bound_spectrum} show two time dependent simulation at two different values of the superconducting phase difference
{\it after} the pulse, $\phi = \phi(t=\infty)$. We see that when the two sets of bound states are
very close in energy the output current beats with a frequency which is given by the level spacing. For well-spaced bound states this frequency is so high that it has no visible effect on the current trace.

\begin{figure}
    \includegraphics[width=0.48\textwidth]{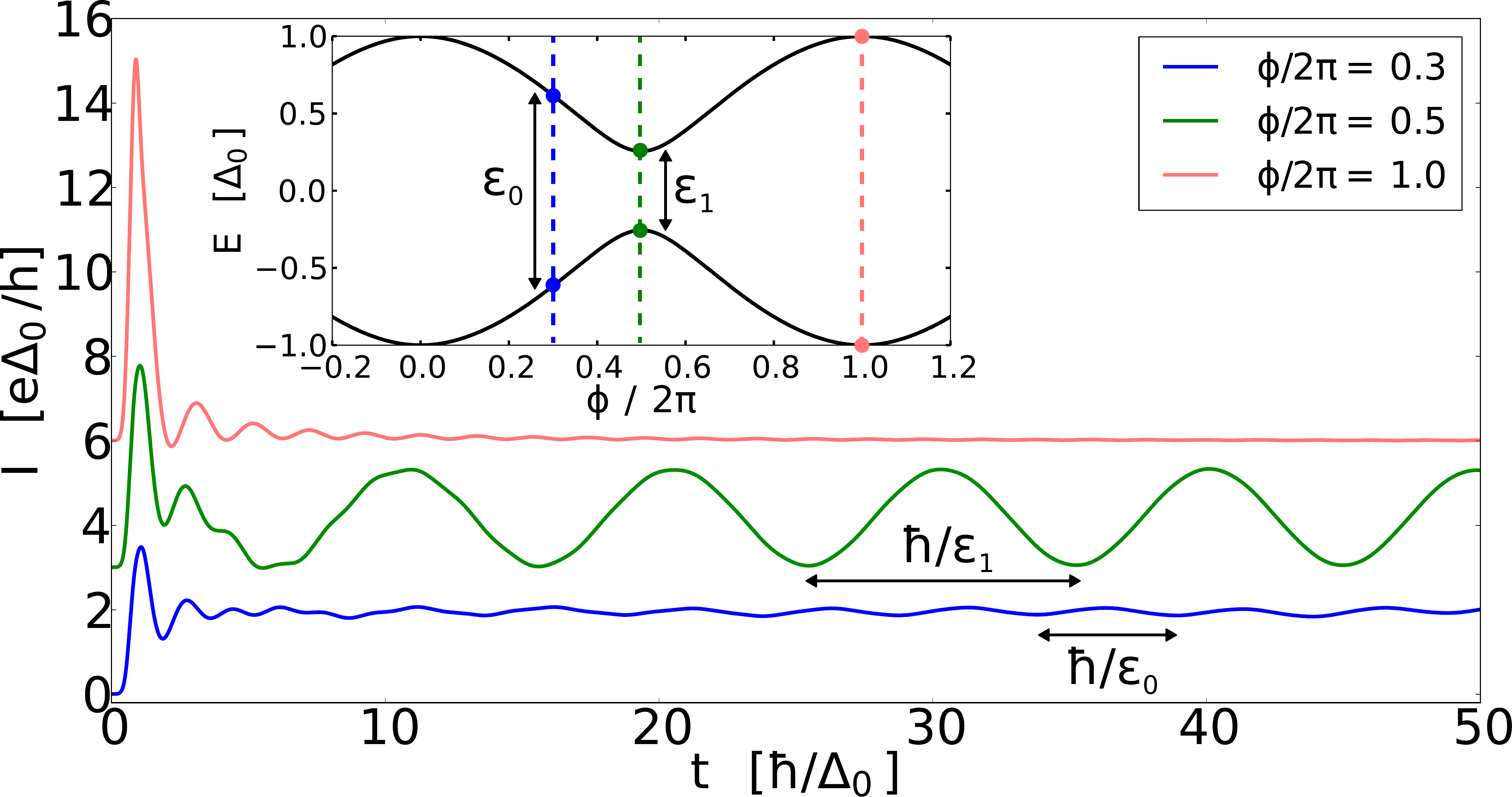}
    \caption{
        Current traces as a function of time for three different voltage pulses
        applied to a short Josephson junction with a transparency of $0.9$. The
        curves have been offset for clarity. Each pulse has a full-width half
        maximum of $0.4\, \hbar/\Delta_0$, and the pulses are of different
        heights. This gives a different phase bias, $\phi$, across the
        junction after the pulse has completed. Inset: The bound state
        spectrum for the junction as a function of the phase bias, the phases
        accumulated by the three pulses are indicated by coloured lines.
    }
    \label{fig:pulse_shortjunction}
\end{figure}

The above effect  is intriguing, but unfortunately long ballistic Josephson junctions are
difficult to realize experimentally (with the exception perhaps of carbon nanotubes). In diffusive junctions
there will be a distribution of times of flight which will wash out the above effect. An alternative is to consider
the limit of short junctions, which have been studied extensively experimentally with atomic contacts (break junctions)\cite{MAR_break_junction}.
We shall, therefore, now explore the effect of a voltage pulse applied to a short Josephson junction.
We do not expect to be able to see a train of well-resolved peaks of current, as in the
long junction case, because the time of flight of the short junction is much shorter than
the typical pulse duration. We do, however, expect to see the effect that gives rise
to the ``beating'' in Fig.~\ref{fig:bound_spectrum}, as this is governed only
by the energy difference between the Andreev bound states in the junction.
Figure~\ref{fig:pulse_shortjunction} shows the current passing through a short
junction when voltage pulses of varying heights are applied. We see an initial
transient part followed by an oscillatory part that continues indefinitely.
Initially, all the states up to $E=0$ are filled (Pauli principle). The pulse
excites some quasiparticles into states at $E>0$ and also shifts the phase
bias across the junction so that we are at a different place in the
phase-energy plot than we were before the pulse (indicated by dashed lines in
the inset to Fig.~\ref{fig:pulse_shortjunction}). Any quasiparticles in continuum
states escape into the leads after some time ($\sim 20 \hbar/\Delta$ in
Fig.~\ref{fig:pulse_shortjunction}), however the contribution in the Andreev
bound states cannot escape. After we have reached steady state we are essentially
in a superposition of Andreev bound states at energy $E$ and $-E$. These two
contributions interfere with one another to give a current that \emph{oscillates}
in time at an angular frequency $2E/\hbar$. This effect is most strongly seen
for $\phi = \pi$, as the Andreev levels have the smallest energy gap here.
For $\phi = 2\pi$ the oscillations die away with time, as the
Andreev levels hybridize with the continuum at this point. By tuning the energy
gap between the Andreev levels after the pulse we are able to control the
frequency of the current. We can tune the energy gap by placing ourselves at
different points in the phase-energy diagram (by sending in pulses of different
heights), or by tuning the transparency of the junction to modify the
phase-energy diagram itself.

The above calculations have been performed in absence of electromagnetic environment.
The closest experimental situation that would correspond to these calculations is a
Josephson junction embedded in a superconducting ring where the voltage pulse is applied through
a pulse of magnetic field through the ring and the signal detected through the magnetization generated
by the oscillating circulating current. A simpler configuration would involve a SQUID where one of the two
junction is an atomic one and the other a regular large tunnel junction.
In a SQUID setup, however, the effect of the electromagnetic environment would have
to be properly included.

\section{Conclusion}
We have developed an algorithm for simulating time-resolved quantum transport,
which we dub ``source-sink'' due to the characteristic addition of
both ``source'' and ``sink'' terms to the Schrödinger-like equations used.
We demonstrated that the accuracy of the method can be tuned at the cost of
increasing the runtime, and that for a given accuracy the algorithm scales
linearly with the system size and the maximum time required. We confirmed the
accuracy of the method by comparing our results for a Josephson junction at
finite bias with analytical results from the literature.

We then studied the effect of a single voltage pulse on a (long or short) Josephson junction.
We found that a single voltage pulse results in a periodic resultant supercurrent.
The (rightly) controversial yet appealing concept of time crystal was recently put forward.\cite{TimeCrystal}
In analogy with a regular crystal where translational spatial symmetry is spontaneously broken,
a time crystal would spontaneously break translational time symmetry. While the above effect is
not a time crystal (the system in the normal part is not in its ground state), it might be as close as one can get;
the superconducting ring remains in its ground state, yet a time dependent current flows through it.

In contrast to other universal effects associated to Josephson physics, the period is
given here by the normal part of the device. In the absence of electromagnetic environment,
the periodic current continue for ever. A precise calculation of the effect of the dissipative
electromagnetic environment to damp the oscillating current is left for future work.

{\it Acknowledgments.}
This work is funded by the ERC consolidator grant
MesoQMC.

\bibliographystyle{apsrev}
\bibliography{references}

\end{document}